\documentclass[apj,iop]{emulateapj}
\slugcomment{The Astrophysical Journal, 753:139 (16pp), 2012 July 10} 
\usepackage[table]{xcolor}
\usepackage[normalem]{ulem}
\usepackage{rotating}
\usepackage{tabularx,booktabs}
\begin{document}

\title{A Multi-epoch Timing and Spectral Study of the Ultraluminous X-ray source \\ NGC 5408 X-1
with \it {XMM-N\lowercase{ewton}}} \author{Dheeraj R. Pasham\altaffilmark{1,2}, Tod E. Storhmayer\altaffilmark{2}}
\affil{$^{1}$Astronomy Department, University of Maryland, College Park, MD 20742; email: dheeraj@astro.umd.edu; richard@astro.umd.edu \\ 
$^{2}$Astrophysics Science Division, NASA's GSFC, Greenbelt, MD 20771; email: tod.strohmayer@nasa.gov \\
{\it Received 2012 January 27; Accepted 2012 May 6; Published 2012 June 22}
}

\begin{abstract}

We present results of new {\it XMM-Newton} observations of the
ultraluminous X-ray source (ULX) NGC 5408 X-1, one of the few ULXs to
show quasi-periodic oscillations (QPO). We detect QPOs in each of four
new ($\approx 100$ ks) pointings, expanding the range of frequencies
observed from 10 - 40 mHz. We compare our results with the timing and
spectral correlations seen in stellar-mass black hole systems, and
find that the qualitative nature of the timing and spectral behavior
of NGC 5408 X-1 is similar to systems in the steep power-law state
exhibiting Type-C QPOs. However, in order for this analogy to
quantitatively hold we must only be seeing the so-called {\it
saturated} portion of the QPO frequency - photon index (or disk flux)
relation.  Assuming this to be the case, we place a lower limit on the
mass of NGC 5408 X-1 of $\ga 800 M_{\odot}$. Alternatively, the QPO
frequency is largely independent of the spectral parameters, in which
case a close analogy with the Type-C QPOs in stellar systems is
problematic.  Measurement of the source's timing properties over a
wider range of energy spectral index is needed to definitively resolve
this ambiguity.  We searched all the available data for both a broad
Fe emission line as well as high frequency QPO analogs (0.1 - 1 Hz),
but detected neither. We place upper limits on the equivalent width of
any Fe emission feature in the 6 - 7 keV band, and of the amplitude
(rms) of a high frequency QPO analog of $\approx 10$ eV and $\approx
4\%$, respectively.

\end{abstract} 


\vfill\eject

\section{Introduction \& Background}

Ultraluminous X-ray sources (ULXs) are bright, point-like, off-nuclear
extragalactic X-ray sources. Since their discovery over two decades
ago (Fabbiano et al. 1989), the true nature of these sources has been
a mystery. Their isotropic X-ray luminosities range from a few
$\times$ $10^{39}$ ergs s$^{-1}$ to as high as $10^{42}$ ergs
s$^{-1}$. Luminosities at the high end of this range are found in a
handful of sources which have been dubbed hyperluminous X-ray sources
(HLXs, e.g., M82 X-1: Matsumoto et al. 2001, Kaaret et al. 2001; HLX
ESO 243-49 X-1: Miniutti et al. 2006, Farrell et al. 2009). The
apparent X-ray luminosities of these sources exceed the Eddington
limit for any known stellar-mass black hole, hence the term
``ultraluminous".

Based on their variability on short timescales (some ULXs are known to
vary on timescales of the order of a few minutes) and high
luminosities, some ULXs are almost certainly powered by accretion onto
black holes (excluding the population of X-ray bright supernovae;
Immler \& Lewin 2003). The controversy, however, is with their mass
range. Black holes, in principle, can exist in three mass
ranges. Stellar-mass black holes (StBH: Mass $\sim$ 3-50 $M_{\odot}$)
are the remnants of stellar evolution of massive normal stars (stars
excluding the POP III category). On the opposite end of the mass scale
are the Super-massive black holes (SMBHs: Mass $\sim 10^{6-10}$
$M_{\odot}$) - which are currently theorized to form either via
accretion onto ``seed'' black holes in the early universe ($\sim$ few
Myrs after the Big Bang) with subsequent mergers (Volonteri et
al. 2003; Davies et al. 2011) or via direct collapse of dense gas
(Begelman et al. 2006). Their exact formation scenario is still
somewhat uncertain. Finally, one has the Intermediate-mass black holes
(IMBHs: Mass range 100 - few $\times$ 1000 $M_{\odot}$), whose
formation mechanisms are the subject of considerable debate. There are
a number of processes whereby an IMBH can form. For example, they can
be formed at the centers of globular clusters via accretion onto a
massive black hole (Mass $\sim$ 50 $M_{\odot}$) over a time period of
$10^{10}$ years (Miller et al. 2002), by accretion onto a primordial
black hole (e.g., Mack et al. 2007), or by mergers of seed black
holes.

The off-centered nature of ULXs rules them out as candidates for
SMBHs. Given the ages of their host galaxies, if they were SMBHs, they
should have sunk to the centers of their galaxies due to dynamical
friction (Miller \& Colbert 2004). Clearly, this is not the case. This
leaves the possibility of their being either StBHs or IMBHs. Theories
supporting the idea of a StBH as the central engine often employ
either beamed X-ray emission or some sort of super-Eddington emission
mechanism to explain the observed high luminosities. Current ideas
include geometric beaming (King et al. 2001), relativistic beaming
similar to the phenomenon of blazars (K\"ording et al. 2002), and
super-Eddington emission as a result of a photon-bubble instability in
radiation-dominated accretion disks (Begelman et al. 2002). More
recently, based on high-quality $XMM$-$Newton$ spectra of 12 ULXs,
Gladstone et al. (2009) have suggested that ULXs might be accreting
StBH systems in a new {\it ultraluminous accretion state} (an addition
to the current classification of black hole accretion states: thermal
(high/soft), hard (low/hard) \& steep power law (very high) as in
McClintock et al. 2006), where accretion is again
super-Eddington. Alternatively, it has been suggested that these
sources might be powered by gravitational energy release from
accretion onto an IMBH (Colbert \& Mushotzky 1999). This presents a
relatively simpler solution, where the energetics proceed similarly to
that of the StBH systems (at sub-Eddington accretion rates) with the
corresponding physical properties scaling appropriately with the mass
of the compact object. It is also possible that ULXs constitute an
inhomogeneous population of both StBHs and IMBHs.

It is now known that certain characteristic timescales in accreting
black hole systems scale with the mass of the black hole. For example,
McHardy et al. (2006) have established that in soft-state StBHs and
AGN, the break timescale of the power density spectrum (PDS), the mass
of the black hole and the luminosity are strongly correlated. Their
work was further extended to the hard-state StBHs by K\"ording et
al. (2007). These studies strongly demonstrate that certain physical
timescales of accreting compact objects scale directly with the mass
of the source (after accounting for differences in accretion rates,
i.e., luminosity). Therefore, timing studies can play a key role in
extracting valuable information about the mass of the compact source, 
%these systems, especially the mass, 
which is crucial in the case of ULXs in order to better understand the
nature of the emission processes that result in such high apparent
luminosities. Of particular interest here are the low frequency
quasi-periodic oscillations (LFQPOs). Within the context of StBHs, the
LFQPOs (those in the range $\sim$ 0.1-15 Hz) are broadly classified
into three categories based on their properties and the overall nature
of their PDS (Casella et al. 2005). The PDS with type-A and B QPOs is
characterized by weak red noise (noise at the low frequency end of the
PDS) with the type-A QPOs occurring with relatively low coherence
(quality factor, Q: centroid frequency/FWHM, $\la$ 3) compared to the
type-B (Q $\ga$ 6). Finally, one has the type-C QPOs which are most
relevant to the present work. The PDS accompanying these QPOs in StBH
systems can be described by a flat-topped, band-limited noise breaking
to a power law with the QPOs evident on the power-law portion of the
spectrum, close to the break.

Further, the type-C QPOs are fairly coherent with the quality factor,
Q from 5-15 and amplitudes (\% rms) ranging from 2-20. In StBHs, they
are known to occur in the frequency range from $\approx$ 0.1-15 Hz. In
light of the work unifying the stellar-mass and the super-massive
black holes, if some ULXs were to host IMBHs then the qualitative
behavior of their PDS should be comparable to StBHs with the
characteristic variability times scaling according to the mass of the
putative IMBH. This idea has been explored by a number of authors to
search for the ``QPO analogs'' in ULXs; and QPOs have now been
detected in a handful of them.
The QPOs detected in M82 X-1 (Strohmayer \& Mushotzky 2003), NGC 5408
X-1 (Strohmayer et al. 2007) and NGC 6946 X-1 (Rao et al. 2010)
resemble the type-C QPOs, while those detected in the M82 source
X42.3+59 more closely resemble the type-A or -B QPOs (Feng et
al. 2010). The crucial difference here is that the QPO centroid
frequencies of the ULX sources appear to be scaled down by a factor of
a few$\times$(10 - 100) ($\sim$ few mHz) compared to the LFQPOs in
StBHs.

Under the assumption that the ULX mHz QPOs are analogs of the LFQPOs
in StBHs, it is reasonable to assume that their characteristic
timescales/frequencies (e.g., QPO centroid frequencies, PDS break
frequencies) scale with the mass of the accreting source and vice
versa. However, the type-C QPOs occur with a wide range of centroid
frequencies ($0.1 - 15$ Hz) in StBH systems. Therefore, timing
information alone is not sufficient to accurately estimate ULX masses
in this way, but combining timing and spectral information has proven
to be a valuable tool. For example, in StBH systems, the power-law
photon index and disk flux are correlated with the QPO centroid
frequency. The general trend is an increase in power-law photon
index and disk flux with the QPO centroid frequency (e.g., Sobczak et
al. 2000; Vignarca et al. 2003; Shaposhnikov \& Titarchuk 2009), with
evidence for saturation (constancy of the power-law photon index and
disk flux with a further increase in QPO centroid frequency) beyond a
certain frequency.

Using a reference StBH system with a measured QPO centroid frequency -
photon-index relation, one can then scale the QPO centroid frequencies
detected in ULX systems at a given power-law spectral index to get an
estimate of the mass. For example, with archival $XMM$-$Newton$ data
from M82 X-1, Dewangan et al. (2006) extracted the energy and power
spectra of the source. The PDS was strikingly reminiscent of a StBH
with type-C QPOs, i.e., flat-top noise breaking to a power-law with
QPOs on the power-law portion of the spectrum. However, the respective
timescales were scaled down by a factor of $\sim$ 10. The QPO in this
case was centered around a frequency of $\sim$ 114 mHz and the energy
spectrum had a power law photon index of $\sim$ 2.0. Using the QPO
frequency - photon index correlations from two StBH reference sources,
GRS 1915+105 and XTE J1550-564, they estimated the mass of the ULX in
M82 X-1 by scaling its QPO centroid frequency ($\sim$ 114 mHz) at the
given photon index ($\sim$ 2.0). They estimated the mass of the black
hole to be in the range 25-520$M_{\odot}$. Similar scaling arguments
were used by Rao et al. (2010) to estimate the mass of the black hole
in the ULX NGC 6946 X-1 to be in the range (1-4) $\times$ 1000
$M_{\odot}$. Based on both the PDS and the energy spectrum of NGC 5408
X-1, Strohmayer et al. (2009) (S09) argued that the source behavior
was consistent with the steep power-law state (SPL) often seen in
StBHs. They compared the available data from NGC 5408 X-1 to five
different StBH reference sources and estimated the mass of the black
hole to be a few $\times$ 1000 $M_{\odot} $. Feng et al. (2010)
detected 3-4 mHz QPOs from the ULX X42.3+59 in M82 and identified them
as either type A/B analogs of StBHs. They estimated the mass of the
black hole to be in the range 12,000-43,000 $M_{\odot}$ by scaling the
QPO frequency to that of the type A/B QPOs in StBHs.

It is important to note that the above scaling arguments have several
caveats. First, the mass estimates were established under the
assumption that the mHz QPOs detected from these ULX systems are
analogous to a particular type of LFQPO detected in StBH systems,
i.e., A, B or C.  These identifications were based on the qualitative
nature of the power spectrum (PDS) alone in some cases and both the
PDS and the energy spectrum in the case of NGC 5408 X-1. Second, in
the case of the ULX mHz QPOs, the observed range of QPO centroid
frequencies and photon indices has been limited. To gain a more secure
identification one would like to see the QPO frequencies and photon
spectral indices {\it correlate} in a similar fashion as for the
StBHs.

%%%%%%%%%%%%%%%%%%%%%%%%%%%%%%%%%%%%%%%%%%%%%%%%%%%
%%%%%%%%%%%%%%%%%%%%%%%%%%%%%%%%%%%%%%%%%%%%%%%%%%%
%%%%%%%%%%%%%%%%%%%%%%%%%%%%%%%%%%%%%%%%%%%%%%%%%%%
\begin{table}
    \caption{{ Summary of the {\it XMM-Newton} observations of NGC 5408 X-1}}\label{Table1} 
{\scriptsize
\begin{center}
   \begin{tabular}[t]{lcccc}
	\hline\hline \\
ObsID & T$_{obs}$\tablenotemark{1} & Net Countrate\tablenotemark{2}& Effective  \\
	& (ks)			   & (counts s$^{-1}$)             & Exposure\tablenotemark{3}(ks) \\
	\\
    \hline \\
 0302900101 (2006) & 132.25 & $1.26 \pm 0.04$  & 99.94  \\

	\\

 0500750101 (2008) & 115.69 & $1.19 \pm 0.04$  & 48.55  \\

	\\

 0653380201 (2010A) & 128.91 & $1.46 \pm 0.04$ & 80.68 \\

	\\

 0653380301 (2010B) & 130.88 & $1.40 \pm 0.04$ & 110.0 \\

	\\

 0653380401 (2011A) & 121.02 & $1.34 \pm 0.07$ & 107.75 \\

	\\

0653380501 (2011B) & 126.37 & $1.31 \pm 0.07$ & 98.66 \\
\\
    \hline\hline
    \end{tabular}
\end{center}
}
\tablenotemark{1}{The total observation time.}\\
\tablenotemark{2}{The average {\it pn+MOS} countrate.}\\
\tablenotemark{3}{After accounting for flaring background and good time intervals. See \S 2 and \S 3 for details on filtering.}\\
\end{table}

%%%%%%%%%%%%%%%%%%%%%%%%%%%%%%%%%%%%%%%%%%%%%%%%%%%
%%%%%%%%%%%%%%%%%%%%%%%%%%%%%%%%%%%%%%%%%%%%%%%%%%%
%%%%%%%%%%%%%%%%%%%%%%%%%%%%%%%%%%%%%%%%%%%%%%%%%%%
In this paper we further explore these issues using extensive new
observations of NGC 5408 X-1 with {\it XMM-Newton}. We describe the
properties of the mHz QPOs over a wider range of centroid frequencies
than previous data allowed. More specifically, we study the nature of
the mHz QPOs from NGC 5408 X-1 through a systematic search for timing
- spectral correlations similar to those seen in the StBHs.  The
article is arranged as follows. In \S 2 we describe the data used for
the present study. In \S 3 we show the results from our timing
analysis, while in \S 4 we give details of the energy spectral
analysis.  In \S 5 we describe our search for timing - spectral
correlations similar to those seen in accreting StBH systems. The
search is conducted using two different spectral models, a
phenomenological model of a $multi$-$colored$ $disk$ + $power$-$law$
and then with a model describing Comptonization by bulk motion
(Titarchuk et al. 1997) that has been used previously to derive black
hole masses from QPO scaling arguments (Shaposhnikov \& Titarchuk
2009). Finally, in \S 6 we discuss the implications of our results on
the mass of the black hole in NGC 5408 X-1.

\section{$XMM$-$Newton$ Observations}
{\it XMM-Newton} has now observed NGC 5408 X-1 on multiple
occasions. We use data from six of the most recent observations
($\sim$ 100 ks each) spread over a time span of five years
(2006-2011). The details of the observations are outlined in Table
1. Results from the first two observations (in 2006 and 2008,
respectively) were summarized in Strohmayer et al. (2007) (S07) \&
Strohmayer et al. (2009) (S09). Using the 2006 data, S07 reported the
first detection of quasi-periodic variability from this source. S09's
analysis of the 2008 observation again showed evidence for the
presence of quasi-periodic oscillations. However, the most prominent
QPO during the 2008 observation was at a lower frequency (QPO at
$\sim$ 10 mHz compared to 20 mHz in 2006). Further, S09 noted that the
disk contribution to the total flux and the power-law index of the
energy spectrum decreased slightly compared to its state in 2006. This
is analogous to a trend often seen in StBH systems where the disk flux
and the power law index of the energy spectrum positively correlate
with the centroid frequency of the most dominant QPO (e.g., Viagnarca
et al. 2003, Sobczak et al. 2000). These findings were used to propose
additional observations with {\it XMM-Newton}, with the goal of
detecting QPOs over a range of frequencies and hence to further
explore for correlations between timing and energy spectral
properties. A large program was approved for Cycle 9 (PI: Strohmayer),
and four observations ($\sim$ 100 ks each) were made under this
program. Two of the observations were carried out in 2010 (2010a \&
2010b) and the rest in 2011 (2011a \& 2011b). Here we present results
from these new observations as well as a reanalysis of the earlier
pointings, so as to facilitate a consistent comparison of all the
available data.

%%%%%%%%%%%%%%%%%%%%%%%%%%%%%%%%%%%%%%%%%%%%%%%%%%%%%%%%%%%%%%%%%

\begin{figure*}
\begin{center}
\includegraphics[width=6.25in, height=7.025in, angle=0]{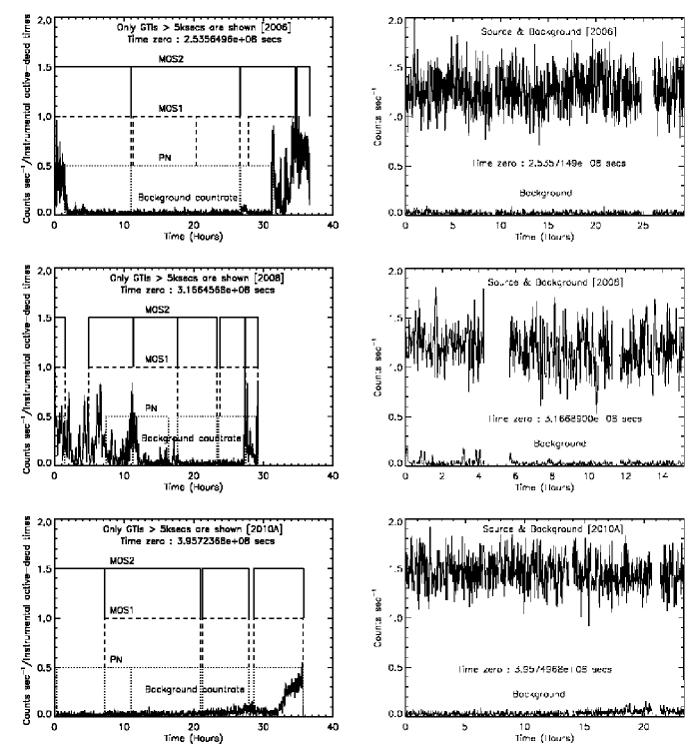}
\end{center}
{\small {\textbf{Figure 1A.} Left Panels: Good time intervals ($>$5 ks)
from EPIC-pn, MOS1 \& 2 onboard $XMM$-$Newton$. For a given
instrument, at a given time (X-axis), a finite value on the y-axis
implies the instrument was \textit{continuously} active for at least 5
ks. Also shown is a combined background (pn+MOS1+MOS2) light curve
with flaring evident in some cases. The observation year is indicated
at the top of each plot (Shown here are {\bf 2006, 2008 \&
2010A data}: see Table 1). Time zero is indicated in secs since 1998.0
TT. Right Panels: Filtered light curves (pn+MOS1+MOS2)
accounting for both instrumental dead times and flaring. The power
spectra were derived from these cleaned light curves. These light
curves were extracted using photons in the energy range of 0.3-10.0
keV. }}
\label{fig:figure1a}
\end{figure*}

%%%%%%%%%%%%%%%%%%%%%%%%%%%%%%%%%%%%%%%%%%%%%%%%%%%

For the present work, we use the data acquired by the European photon
imaging camera (EPIC), i.e, both the {\it pn} and {\it MOS} to get a
higher signal-to-noise ratio in the power and energy spectra. We used
the standard SAS version 11.0.0 to reduce the images and filtered
event-lists from all the EPIC data. The standard filter with {\it
(PATTERN$<$=4)}, to include only single and double pixel events, was
applied to the event-lists and events only in the energy range
0.3-12.0 keV were considered for further analysis. Background flaring
was prominent for brief periods during certain observations. The power
and energy spectra were carefully extracted taking into account both
the background flaring and the instrumental dead time effects
(specific details in the next sections). In all the observations the
source was easily detectable and we did not face any source confusion
problems. We extracted source events from a region of radius 32''
centered around the source. This particular value was chosen to
roughly include 90\% of the light from the source (estimated from the
fractional encircled energy of the EPIC instruments). A background
region, free of other sources, was extracted in a nearby region. The
size of the background region was chosen to be consistent with the
source region. Further, the size of the source and the background
region was chosen to be consistent for all the six observations (i.e.,
32''). We present the specific details of the timing and spectral
analysis in the following sections.

%%%%%%%%%%%%%%%%%%%%%%%%%%%%%%%%%%%%%%%%%%%%%%%%%%%%%%%%%%%%%%%%%

\begin{figure*}
\begin{center}
\includegraphics[width=6.25in, height=7.025in, angle=0]{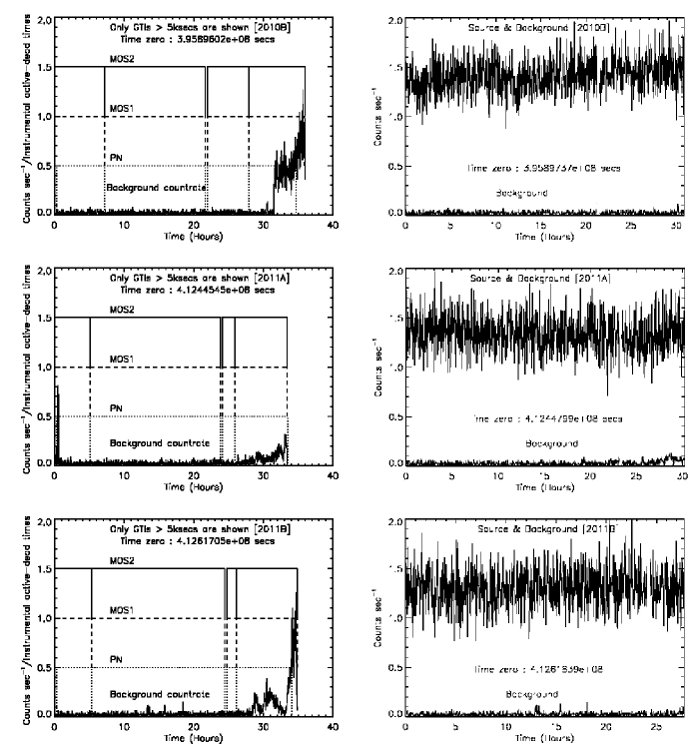}
\end{center}
{\small{\textbf{Figure 1B.} Left Panels: Good time intervals ($>$5 ks)
from EPIC-pn, MOS1 \& 2 onboard $XMM$-$Newton$. For a given
instrument, at a given time (X-axis), a finite value on the y-axis
implies the instrument was \textit{continuously} active for at least 5
ks. Also shown is a combined background (pn+MOS1+MOS2) light curve
with flaring evident in some cases. The observation year is indicated
at the top of each plot (Shown here are {\bf 2010B, 2011A \& 2011B
data}: Refer to Table 1). Time zero is indicated in secs since 1998.0
TT. Right Panels: Filtered light curves (pn+MOS1+MOS2)
accounting for both instrumental dead times and flaring. The power
spectra were derived from these cleaned light curves. These light
curves were extracted using photons in the energy range of 0.3-10.0
keV.}}
\label{fig:figure1b}
\end{figure*}
%%%%%%%%%%%%%%%%%%%%%%%%%%%%%%%%%%%%%%%%%%%%%%%%%%%

\section{Timing Analysis}

We first produced the source and the background light curves to assess
the quality of the data. Shown in the left panels of Figure 1A and 1B
are the background count rates ({\it pn+MOS1+MOS2}) for each of the
six observations. Overlaid are the time intervals during which a given
instrument ({\it pn, MOS1/2}) was continuously active for more than 5
ks, i.e., all the instrumental good time intervals (GTIs) longer than
5 ks. For a given instrument, a horizontal line (offset to an
arbitrary value for each instrument) indicates the active time, while
a vertical line marks the beginning or the end of a continuous data
interval. Such an insight is important as we are using both the {\it
pn} and the {\it MOS} data to achieve higher count rates compared to,
say, the {\it pn} alone. In other words, this ensures that the
combined {\it pn} and {\it MOS} event-lists contain only events
corresponding to the times during which all the three instruments were
active. Furthermore, inspection of the light curves reveals background
flaring at either the beginning or the end, or both, of all the
observations. Therefore, the final combined {\it (pn+MOS1+MOS2)} light
curves were extracted by taking into account the GTIs of all the
instruments and excluding the periods of background flaring. The
filtered light curves (100s bins) of the source and the non-flaring
background in the energy range 0.3 - 12.0 keV are shown in the right
panels of Figures 1A and 1B. The combined mean count rates (0.3-12.0
keV) during each observation are listed in Table 1. Notice that the
gaps in the data are either due to exclusion of background flaring
times, dead time intervals or the absence of GTIs longer than 5 ks.

%%%%%%%%%%%%%%%%%%%%%%%%%%%%%%%%%%%%%%%%%%%%%%%%%%%
%%%%%%%%%%%%%%%%%%%%%%%%%%%%%%%%%%%%%%%%%%%%%%%%%%%
%%%%%%%%%%%%%%%%%%%%%%%%%%%%%%%%%%%%%%%%%%%%%%%%%%%

%\oddsidemargin=-0.9cm
%\vspace{-0.5cm}
\begin{table*}
  %\begin{flushleft}
  \caption{{Summary of the best-fitting model parameters for the power density spectra (PDS) of the ULX NGC 5408 X-1. Here, we use an energy band in which the QPO is detected with very high confidence.}}\label{Table2} 
\centering
%\vspace{-0.5cm}
{\scriptsize
    \begin{tabular}[t]{lccccccc}
     %\begin{tabular*}{20cm}[t]{lccccccc}
    \hline\hline \\
   Dataset & 2006 & 2008 & 2010A & 2010B & 2011A & 2011B \\
	\\
    \hline \\
Exposure\tablenotemark{a}(ks)         & 27.6$\times$3 & 15.5$\times$2 & 23.2$\times$3 & 21.0$\times$4 & 67.4$\times$1 & 55.0$\times$1\\
\\
\rowcolor[gray]{.8} Energy Range\tablenotemark{b}(keV)    & 0.30-2.0 & 1.0-10.0 & 0.85-10.0 & 1.0-4.0 & 1.10-8.0 & 1.20-2.0 \\
\\
Countrate\tablenotemark{c}(cts s$^{-1}$) & $1.17 \pm 0.03$ & $0.43 \pm 0.02$ & $0.71 \pm 0.02$ & $0.50 \pm 0.01$ & $0.41 \pm 0.02$ & $0.23 \pm 0.01$ \\
\\
C\tablenotemark{$\ast$}	& $1.95 \pm 0.01$ & $1.81 \pm 0.02$ & $1.95 \pm 0.02$ & $2.00 \pm 0.01$ &  $1.95 \pm 0.02$ & $1.92 \pm 0.02$ \\ 
\\
A\tablenotemark{$\ast$}	& $0.86 \pm 1.05$ & $3.39 \pm 2.90$ & $9.14 \pm 17.90$ & $(0.80 \pm 2.30)\times10^{-04}$ &  $3.49 \pm 10.61$ & $(2.11 \pm 3.47)\times10^{-03}$ \\   
\\
$\Gamma_{Low}$\tablenotemark{$\ast$}		& $0.20 \pm 0.19$ & $0.09 \pm 0.13$ & $-0.25 \pm 0.31$ & $2.68 \pm 0.98$ &  $-0.04 \pm 0.42$ & $1.45 \pm 0.52$ \\ 
\\
$\nu_{bend}$\tablenotemark{$\ast$}(mHz)	& $6.33 \pm 0.75$ & $6.34 \pm 0.49$ & $10.26 \pm 6.03$ & $30.63 \pm 20.20$ &  $5.80 \pm 4.84$ & $7.15 \pm 8.67$ \\ 
\\
$\Gamma_{High}$\tablenotemark{$\ast$}		& $6.74 \pm 3.34$ & $10.25 \pm 5.60$ & $1.60 \pm 0.50$  & $0.25 \pm 0.17$ &  $1.98 \pm 1.83$ & $-0.09 \pm 0.43$ \\ 
\\
N$_{QPO,1}$\tablenotemark{$\dagger$}		& $1.88 \pm 0.16$ & $3.51 \pm 1.69$ & $0.63 \pm 0.15$ & $0.72 \pm 0.21$ &  $2.10 \pm 0.56$ & $1.04 \pm 0.29$ \\
\\
$\nu_{0,1}$\tablenotemark{$\dagger$}(mHz) 	& $17.68 \pm 0.77$ & $10.28 \pm 0.24$ & $40.40 \pm 2.93$ & $38.81 \pm 2.01$ &  $18.67 \pm 2.20$ & $19.43 \pm 0.90$ \\  
\\
$\Delta\nu_{1}$\tablenotemark{$\dagger$}(mHz)	& $15.48 \pm 2.05$ & $1.28 \pm 0.97$ & $28.06 \pm 11.96$ & $18.75 \pm 8.98$ &  $21.98 \pm 6.23$ & $7.08 \pm 3.60$ \\
\\
N$_{QPO,2}$\tablenotemark{$\dagger$}		& - & $1.91 \pm 0.35$ & - & - & - & -  \\
\\
$\nu_{0,2}$\tablenotemark{$\dagger$}(mHz)	& - & $15.05 \pm 1.68$ & - & - & - & -  \\
\\
$\Delta\nu_{1}$\tablenotemark{$\dagger$}(mHz)	& - & $11.10 \pm 3.05$ & - & - & - & -  \\
\\
\hline
\\
$\chi^2$/dof            & 309.83/208    & 753.75/764    & 178.14/137    & 128.05/97   & 327.11/329 & 268.86/267 \\
(continuum\tablenotemark{d})		& (618.71/211)  & (868.30/770)  & (257.33/140) & (207.54/100) & (669.58/332) & (543.03/270) \\
%\\
%Confidence level             & 0.55  & 0.82  & 0.85  & 0.85 & 1.1 & 1.4  \\
\\
    \hline\hline
    \end{tabular}
}
\\
  %\end{flushleft}
%\vspace{-0.25cm}
\begin{flushleft}
\tablenotemark{a}{The good time intervals were broken into smaller intervals to improve the signal-to-noise in the power density spectra (PDS). For a given observation, the size of each segment$\times$number of such segments is shown.}\\
\tablenotemark{b}{\bf The power spectrum was derived using all the photons in this energy range. For a given observation, this is an energy range in which the QPO was detected with a high significance.} \\
\tablenotemark{c}{The countrate in the bandpass shown in the second row.}\\
\tablenotemark{$\ast$}{We fit the continuum with a bending power-law model described as follows: \\
\begin{center}
\begin{math} Continuum = C + \frac {A\nu^{-\Gamma_{Low}} } {1 + \left(\frac {\nu} {\nu_{bend}}\right)^{\Gamma_{High}-\Gamma_{Low}} }\end{math}
\end{center}
where, $\Gamma_{Low}$ and $\Gamma_{High}$ are the low and high frequency slopes, respectively, and $\nu_{bend}$ is the bend frequency.
}\\
\tablenotemark{$\dagger$}{We model the QPOs with a Lorentzian. The functional form is as follows: 
\begin{center}
\begin{math} QPO = \frac {N_{QPO}} {1 + \left(\frac {2(\nu - \nu_{0})} {\Delta\nu_{0}}\right)^{2} } \end{math}
\end{center}
where, $\nu_{0}$ is the centroid frequency and $\Delta$$\nu_{0}$ is the FWHM of the QPO feature.\\
\tablenotemark{d}{The $\chi^2$/dof for the continuum are shown in braces.}\\
}
\end{flushleft}
\end{table*}

%%%%%%%%%%%%%%%%%%%%%%%%%%%%%%%%%%%%%%%%%%%%%%%%%%%

It is evident from the light curves (right panels of Figure 1A and 1B)
that the source varies significantly during all the observations. To
quantify the variability, we construct a power density spectrum
from each of the light curves. To achieve higher signal-to-noise in
the PDS, we break the GTIs into shorter segments (S/N $\propto$
$\surd$Number of individual spectra, e.g., Van der klis 1989) and
derive an average power spectrum. For a given observation, the size
and the number of such segments are shown in the first row of Table 2.
Figures 2A and 2B show the PDS and the best-fitting model {\it (thick
solid line)} for each of the six observations in two different energy
bands. Shown in the left panel is a PDS in an energy band in which a
QPO is detected with high statistical significance and in the right
panel is a PDS derived from photons in the energy band of 1.0-10.0 keV
(the reason for constructing power spectra in two different energy
bands is discussed in the next paragraph). All the power spectra shown
here use the so-called Leahy normalization, with the Poisson noise
level being 2 (Leahy et al. 1983). Clearly, in each spectrum the power
rises below $\sim$ 0.1 Hz with evidence for a QPO in the range of
10-40 mHz and essentially Poisson noise at higher frequencies.

%%%%%%%%%%%%%%%%%%%%%%%%%%%%%%%%%%%%%%%%%%%%%%%%%%%
%%%%%%%%%%%%%%%%%%%%%%%%%%%%%%%%%%%%%%%%%%%%%%%%%%%

\begin{figure*}
\begin{center}
\includegraphics[width=6.25in, height=6.75in, angle=0]{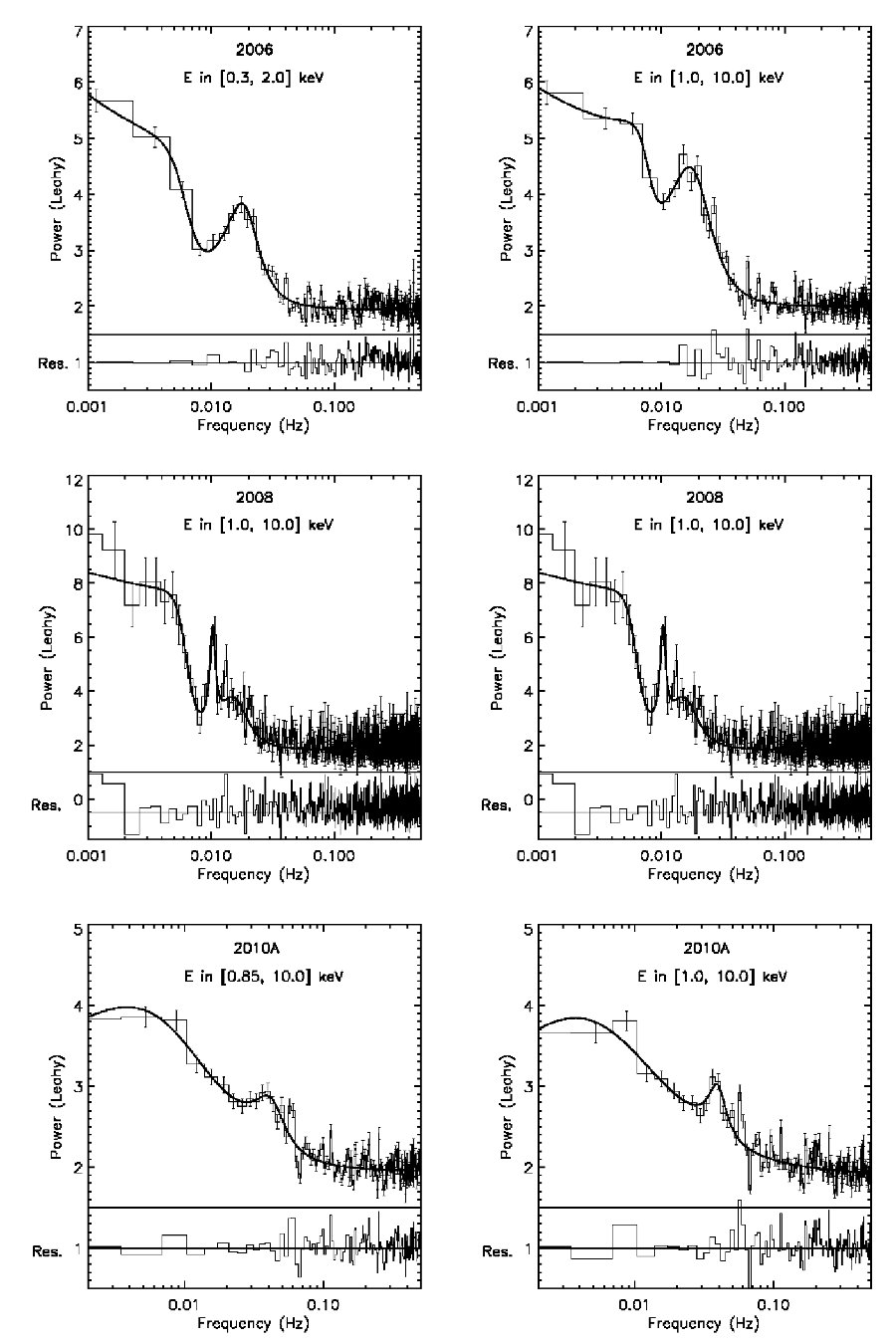}
\end{center}
{\small {\textbf{Figure 2A:} Shown are the power spectra derived from
the first three of the six {\it XMM-Newton} observations ({\bf 2006,
2008, 2010A}: Table 1). For a given power spectrum the strength of a
QPO is dependent on the energy band under consideration. Left
Panels: The power spectrum using an energy band (shown at the top of
each plot) in which the QPO is clearly detected. Also shown are the
 error bars and the residuals (Data-Model) offset to an arbitrary value in each case. Right Panels: Same as the plots on the left panel. However, here we
choose a fixed energy range of 1.0-10.0 keV. A consistent energy band
across all the power spectra is required to unambiguously assess the
behavior (especially, rms amplitude versus QPO centroid frequency:
Fig.3) of these QPOs.}}
\label{fig:figure1a}
\end{figure*}

%%%%%%%%%%%%%%%%%%%%%%%%%%%%%%%%%%%%%%%%%%%%%%%%%%%
%%%%%%%%%%%%%%%%%%%%%%%%%%%%%%%%%%%%%%%%%%%%%%%%%%%

%%%%%%%%%%%%%%%%%%%%%%%%%%%%%%%%%%%%%%%%%%%%%%%%%%%

\begin{figure*}
\begin{center}
\includegraphics[width=6.25in, height=6.75in, angle=0]{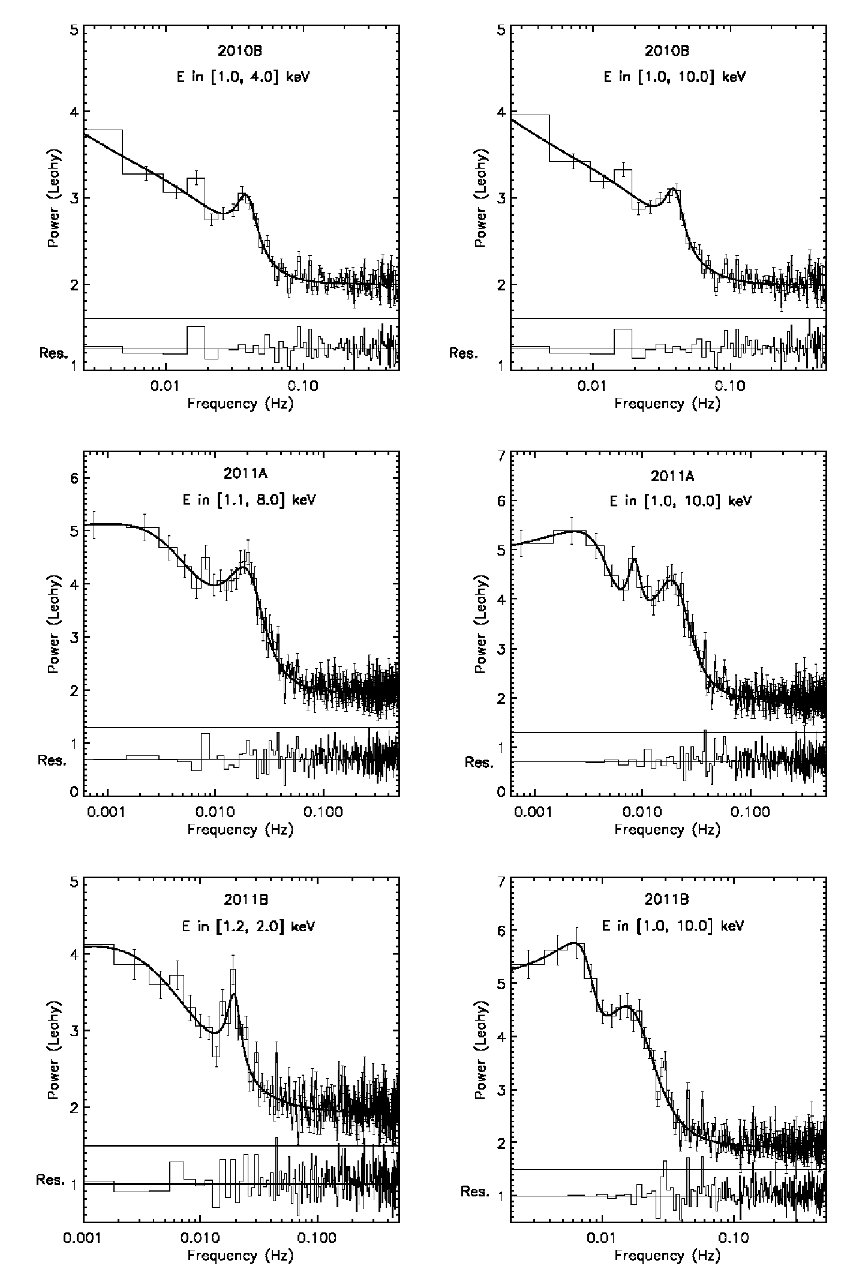}
\end{center}

{\small {\textbf{Figure 2B:} Shown are the power spectra derived from
the last three of the six {\it XMM-Newton} observations ({\bf 2010B,
2011A, 2011B}: Table 1). For a given power spectrum, the strength of a
QPO is dependent on the energy band under consideration. Left
Panels: The power spectrum using an energy band (shown at the top of
each plot) in which the QPO is clearly detected. Also shown are the
 error bars and the residuals (Data-Model) offset to an arbitrary value in each case. Right Panels: Same as the plots on the left panel. However, here we
choose a fixed energy range of 1.0-10.0 keV. A consistent energy band
across all the power spectra is required to unambiguously assess the
behavior (especially, rms amplitude versus QPO centroid frequency:
Fig.3) of these QPOs.}}
\end{figure*}

%%%%%%%%%%%%%%%%%%%%%%%%%%%%%%%%%%%%%%%%%%%%%%%%%%%
%%%%%%%%%%%%%%%%%%%%%%%%%%%%%%%%%%%%%%%%%%%%%%%%%%%

To quantify this behavior, we fit a bending power law to the continuum
and a Lorentzian to model the QPO (Belloni et al. 2002). The model
fits well with the reduced $\chi^{2}$ $\approx$ 1 (last row of Tables
2 \& 3) in three cases (2008, 2011A, 2011B) and gives acceptable fits
with the reduced $\chi^{2}$ in the range 1.3-1.5 (last row of Tables 2
\& 3) in the other three cases (2006, 2010A, 2010B). A careful
analysis of the residuals in the latter three cases indicates that
multiple weak features contribute significantly to increase the
overall $\chi^{2}$. For example in the case of the PDS derived from
the 2010A data (bottom two panels of Figure 2A), the weak QPO-like
feature at $\sim$ 55 mHz and the excess at $\sim$ 0.2 Hz contribute
about 30 to the total $\chi^{2}$. However, their individual
statistical significance is rather low. Nevertheless, for the purposes
of analyzing the QPO properties and studying the overall qualitative
nature of the power spectra, the fits are adequate. The best-fitting
model parameters (derived from a fit in the frequency range 1.0 mHz -
0.5 Hz) for each observation are shown in Tables 2 \& 3 for the
two different energy bands (highlighted in the tables). Also shown are
the $\chi^{2}$/degrees of freedom (dof) values for each of the fits
along with the $\chi^{2}$/dof corresponding to the continuum model (in
braces). The change in $\chi^{2}$ serves as an indicator of the
statistical significance of the QPOs. For a given observation, we
choose two different energy bands: the first energy band (second row
of Table 2) is the bandpass in which the QPO is detected with a very
high confidence and the second bandpass of 1.0-10.0 keV was chosen to
consistently compare the properties of the QPO across all the
observations (overall variability is energy dependent). We confirm the
overall qualitative nature of the PDS, that is, flat-topped noise
breaking to a power law with a QPO on the power law portion (near the
break). This is consistent with StBHs showing type-C QPOs but with the
characteristic frequencies scaled down by a factor of a
few$\times$10. This is in agreement with the results reported by S07
and S09, except with regard to possible close QPO pairs, as here all
our PDSs (Figure 2A \& 2B) are averaged to the extent of smearing out
weak features (to clearly identify the strongest QPO). The effect of
averaging is evident from the best-fit parameters of the QPOs (Tables
2 \& 3). More specifically, since we have smeared out the possible
close pair QPOs (as reported by S07 \& S09 using the 2006 and the 2008
datasets, respectively) the average quality factors of the QPOs
reported here are relatively lower than those typical of the type-C
QPOs from StBHs. Moreover, variation of the QPO centroid frequency
over the timescale of the observation can further decrease its
coherence. Furthermore, we analyzed the PDSs of the backgrounds from
each of the six observations and note that they are all consistent
with a constant Poisson noise.

%%%%%%%%%%%%%%%%%%%%%%%%%%%%%%%%%%%%%%%%%%%%%%%%%%%

\begin{figure*}

%\centering
%\vspace{-0.75cm}
\begin{center}
\includegraphics[width=6.5in, height=3.25in, angle=0]{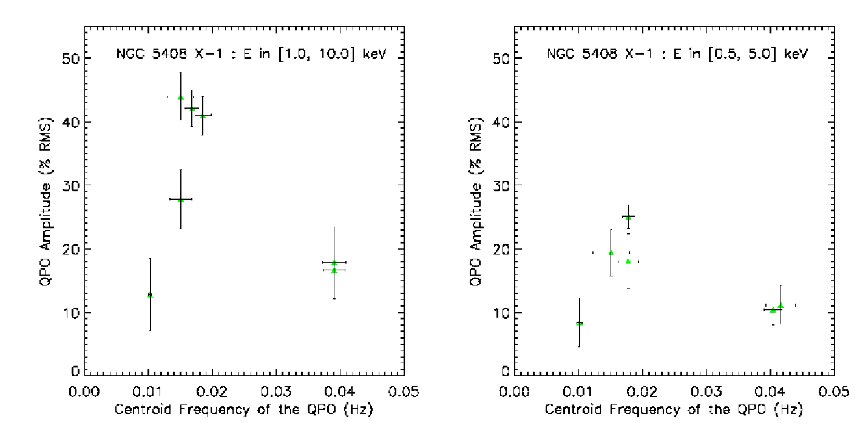}
\end{center}

{\small {\textbf{Figure 3:} The rms amplitude of the mHz QPOs detected from
NGC 5408 X-1 (Y-axis) is plotted against the QPO centroid frequency
(X-axis). To study the dependence of this plot on the bandpass, we
consider two different energy bands. {\it [Left Panel]:} Centroid
frequency versus QPO amplitude using photons in the energy range from
1.0-10.0 keV.  {\it [Right Panel]:} Same as the left panel, but a
different band-pass of 0.5-5.0 keV was used. The qualitative nature of
the behavior does not change with the energy band under
consideration. However, the rms amplitude of a given QPO seems to
increase with increasing energy (at least in the two energy bands
considered), i.e., higher rms amplitude in 1.0-10.0 keV compared to
0.5-5.0 keV.}}
\label{fig:figure1a}
\end{figure*}

%%%%%%%%%%%%%%%%%%%%%%%%%%%%%%%%%%%%%%%%%%%%%%%%%%%
%%%%%%%%%%%%%%%%%%%%%%%%%%%%%%%%%%%%%%%%%%%%%%%%%%%

One of the main goals of the present work is to better characterize
the mHz QPOs seen from NGC 5408 X-1 within the context of the known
classes of LFQPO seen in StBHs (Casella et al. 2005). An important
diagnostic for this purpose is the study of the variation of the rms
amplitudes of these QPOs with their centroid frequency. In the case of
LFQPOs from StBH systems this behavior is fairly well-established
(e.g., Revnivtsev et al. 2000; Sobczak et al. 2000; Vignarca et
al. 2003; McClintock et al. 2006). The typical behavior of the StBH
LFQPOs, in the energy range 2.0-20.0 keV (the nominal RXTE bandpass),
can be described as follows: as the source evolves along the low/hard
state towards the SPL state, the QPO frequency increases and the rms
amplitude increases.  Before reaching the SPL state, the source
traverses an ``intermediate'' state in which the QPO frequency
continues to increase, but the rms amplitude decreases. Upon reaching
the SPL state the correlation tends to break down, showing more
scatter in the rms amplitude (see Figure 11 in McClintock et
al. 2009).
%%%%%%%%%%%%%%%%%%%%%%%%%%%%%%%%%%%%%%%%%%%%%%%%%%%
%%%%%%%%%%%%%%%%%%%%%%%%%%%%%%%%%%%%%%%%%%%%%%%%%%%
%\vspace{-0.5cm}
\begin{table*}
  %\begin{flushleft}
  \caption{{ Summary of the best-fitting model parameters for the power density spectra (PDS) (using only photons in the energy range of 1.0-10.0 keV) of the ULX NGC 5408 X-1.}}\label{Table2} 
\centering
%\vspace{-0.25cm}
{\scriptsize
    \begin{tabular}{lccccccc}
     %\begin{tabular*}{20cm}[t]{lccccccc}
    \hline\hline \\
   Dataset & 2006 & 2008 & 2010A & 2010B & 2011A & 2011B \\
	\\
    \hline \\
Exposure\tablenotemark{a} (ks)        & 27.6$\times$3 & 15.5$\times$2 & 23.2$\times$3 & 21.0$\times$4 & 67.4$\times$1 & 55.0$\times$1\\
\\
\rowcolor[gray]{.8} Energy Range (keV)    & 1.0-10.0 & 1.0-10.0 & 1.0-10.0 & 1.0-10.0 & 1.0-10.0 & 1.0-10.0 \\
\\
Countrate\tablenotemark{b}(cts s$^{-1}$)	& $0.43 \pm 0.01$ & $0.43 \pm 0.02$ & $0.55 \pm 0.02$ & $0.53 \pm 0.01$ &  $0.49 \pm 0.02$ & $0.49 \pm 0.02$ \\
\\
C\tablenotemark{$\ast$}	& $2.00 \pm 0.01$ & $1.81 \pm 0.02$ & $1.90 \pm 0.04$ & $1.99 \pm 0.01$ &  $1.94 \pm 0.01$ & $1.92 \pm 0.01$ \\  
\\
A\tablenotemark{$\ast$}	& $0.80 \pm 0.89$ & $3.39 \pm 2.90$ & $0.01 \pm 0.01$ & $(1.48  \pm 3.88)\times10^{-04}$ &  $4.16 \pm 11.49$ & $2.96 \pm 9.50$ \\
\\
$\Gamma_{Low}$\tablenotemark{$\ast$}		& $0.20 \pm 0.17$ & $0.09 \pm 0.13$ & $1.12 \pm 0.30$ & $2.65 \pm 1.06$ &  $-0.072 \pm 0.39$ & $-0.05 \pm 0.56$ \\
\\
$\nu_{bend}$\tablenotemark{$\ast$}(mHz)	& $7.89 \pm 0.73$ & $6.34 \pm 0.49$ & $7.12 \pm 6.13$   & $36.39 \pm 21.26$ &  $4.96 \pm 1.48$ & $8.30 \pm 1.15$ \\
\\
$\Gamma_{High}$\tablenotemark{$\ast$}		& $10.02 \pm 11.47$ & $10.25 \pm 5.60$ & $-0.41 \pm 0.46$ & $0.25 \pm 0.14$ &  $4.42 \pm 7.78$ & $9.87 \pm 12.11$ \\  
\\
N$_{QPO,1}$\tablenotemark{$\dagger$}		& $2.49 \pm 0.18$ & $3.51 \pm 1.69$ & $0.64 \pm 0.15$ & $0.67 \pm 0.21$ &  $2.40 \pm 0.19$ & $2.64 \pm 0.27$ \\ 
\\
$\nu_{0,1}$\tablenotemark{$\dagger$}(mHz) 	& $16.81 \pm 1.05$ & $10.28 \pm 0.24$ & $39.07 \pm 1.70$ & $39.07 \pm 1.82$ &  $18.49 \pm 1.28$ & $15.04 \pm 2.04$ \\
\\
$\Delta\nu_{1}$\tablenotemark{$\dagger$}(mHz)	& $19.51 \pm 2.16$ & $1.28 \pm 0.97$ & $15.15 \pm 6.44$ & $16.04 \pm 8.89$ &  $21.86 \pm 2.60$ & $22.83 \pm 2.94$ \\ 
\\
\rowcolor[gray]{.95} RMS$_{QPO,1}$\tablenotemark{$\dagger$}           & $42.12 \pm 2.83$ & $12.81 \pm 5.75$ & $16.64 \pm 4.05$ & $17.84 \pm 5.68$ &  $41.00 \pm 3.05$ & $43.94 \pm 3.72$ \\ 
\\
N$_{QPO,2}$\tablenotemark{$\dagger$}		& -	& $1.91 \pm 0.35$ & - &	- & $1.31 \pm 1.02$ & -  \\
\\
$\nu_{0,2}$\tablenotemark{$\dagger$}(mHz)	        & -	& $15.05 \pm 1.68$ & -	& - & $8.40 \pm 1.22$	&  -\\
\\
$\Delta\nu_{1}$\tablenotemark{$\dagger$}(mHz)	& -	& $11.10 \pm 3.05$ & -	& - & - &  -\\
\\
\rowcolor[gray]{.95} RMS$_{QPO,2}$\tablenotemark{$\dagger$}           & - & $27.82 \pm 4.64$ & - & - & $3.09 \pm 4.32$ & - \\ 
\\
\hline
\\
$\chi^2$/dof            & 268.34/208    & 753.75/764    & 200.60/137    & 135.27/97   & 350.16/326 & 291.45/266 \\
(continuum\tablenotemark{d})		& (755.28/211)  & (868.30/770)  & (242.10/140) & (190.64/100) & (851.45/332) & (677.00/269) \\
%\\
%Confidence level             & 0.55  & 0.82  & 0.85  & 0.85 & 1.1 & 1.4  \\
\\
    \hline\hline
    \end{tabular}
}\\
%\vspace{.5in}
\begin{flushleft}
\tablenotemark{a}{The good time intervals were broken into smaller intervals to improve the signal-to-noise in the power density spectra (PDS). For a given observation, the size of each segment$\times$number of such segments is shown. Same time intervals as in Table 1 were used here.}\\
\tablenotemark{b}{\bf To systematically study the QPO properties, we reduced the power spectra in a consistent energy band of 1.0-10.0 keV. } \\
\tablenotemark{c}{The countrate in the bandpass of 1.0-10.0 keV.}\\
\tablenotemark{$\ast$}{We fit the continuum with a bending power-law model described as follows: \\
\vspace{-0.1cm}
\begin{center}
\begin{math} Continuum = C + \frac {A\nu^{-\Gamma_{Low}} } {1 + \left(\frac {\nu} {\nu_{bend}}\right)^{\Gamma_{High}-\Gamma_{Low}} }\end{math}
\end{center}
\vspace{-0.1cm}
where, $\Gamma_{Low}$ and $\Gamma_{High}$ are the low and high frequency slopes, respectively, and $\nu_{bend}$ is the bend frequency. 
}\\
\tablenotemark{$\dagger$}{We model the QPOs with a Lorentzian. The functional form is as follows: 
\begin{center}
\begin{math} QPO = \frac {N_{QPO}} {1 + \left(\frac {2(\nu - \nu_{0})} {\Delta\nu_{0}}\right)^{2} }, \hspace{15pt} RMS_{QPO} (integrated \hspace{2pt} from \hspace{2pt} -\infty \hspace{2pt} to \hspace{2pt} +\infty) = \frac {\pi N_{QPO}\Delta\nu} {2}\times100 \end{math}
\end{center}
where, $\nu_{0}$ is the centroid frequency and $\Delta$$\nu_{0}$ is the FWHM of the QPO feature.\\
\tablenotemark{d}{The $\chi^2$/dof for the continuum are shown in braces.}\\
}
\end{flushleft}
\end{table*}

%%%%%%%%%%%%%%%%%%%%%%%%%%%%%%%%%%%%%%%%%%%%%%%%%%%
The X-ray variability of NGC 5408 X-1 is known to depend on energy
(Middleton et al. 2011). The overall rms variability increases with an
increase in the energy of the photons at least up to 2.5 keV (Given
the poor signal-to-noise at higher energies, it is not clear whether
the variability strength levels off or decreases). This is similar to
the energy dependence of LFQPO ($\approx$ few Hz) detected from the
galactic micro-quasar GRS 1915+105 (Rodriguez et al. 2004), where the
QPO amplitude increases with energy, before rolling over.  We used the
QPO fit parameters from all the observations to explore the dependence
of the QPO rms amplitude with centroid frequency.  To be consistent
across all the observations, we chose an optimum bandpass of 1.0-10.0
keV to ensure good statistics and significant detection of the
variability (the variability of the source increases at higher
energies while the countrate decreases). In addition, to explore
whether our results might depend on energy, we derived the QPO
properties in a different energy band (though with some overlap of the
optimum band defined above) of 0.5-5.0 keV. The rms amplitude versus
QPO frequency results for the 1.0-10.0 keV and 0.5-5.0 keV are shown
in the left and the right panels of Figure 3, respectively (compare
with Figure 11 of McClintock et al. 2009).  Given that the QPO
amplitude is higher at higher energies (Middleton et al. 2011), not
all power spectra derived from the lower energy band (0.5 - 5.0 keV)
yielded statistically significant QPO features. This is why the right
panel of Figure 3 has one less data point.  Clearly, the variation is
similar in both the energy bands. We further compare with the results
from StBHs in \S 5.

Another prime driver for the long observations of NGC 5408 X-1 was to
search for high frequency QPO (HFQPO) analogs of StBHs. These QPOs are
observed in the range from about 50 - 450 Hz in StBHs (McClintock \&
Remillard 2006; Strohmayer 2001a). Simple mass scaling arguments would
suggest that in IMBHs they would be expected in the frequency range of
0.1-1 Hz, with a low rms amplitude of $\sim$ 2\%. We do not detect any
obvious QPO-like feature at these frequencies in any of the individual
observations. Further, given that HFQPOs appear to be reasonably
stable in frequency in StBHs (e.g., Strohmayer 2001b), we averaged the
power spectra derived from the individual observations to improve the
signal-to-noise, however, this did not lead to a detection. We
estimate the upper limit on the rms amplitude of a QPO-like feature in
the frequency range of 0.1-1.0 Hz to be $\sim$ 4\%.

%%%%%%%%%%%%%%%%%%%%%%%%%%%%%%%%%%%%%%%%%%%%%%%%%%%
%%%%%%%%%%%%%%%%%%%%%%%%%%%%%%%%%%%%%%%%%%%%%%%%%%%

%\oddsidemargin=-0.9cm
%\vspace{-0.75cm}
\begin{table*}
%\oddsidemargin=-1.9cm
  %\begin{flushleft}
  \caption{{ Summary of the energy spectral modeling of NGC 5408 X-1. {\it [Top Panel]:} Best-fitting parameters for the {\it tbabs$\ast$(diskpn+apec+pow)} model}. {\it [Bottom Panel]:} Best-fitting parameters for the {\it tbabs$\ast$(bmc+apec)} model. }\label{Table4} \centering
{\scriptsize
    \begin{tabular}{lcccccccc}
     %\begin{tabular*}{20cm}[t]{lccccccc}
    \hline\hline \\	
& & & & {\it tbabs$\ast$(diskpn+apec+pow)} & & & \\
\\
    \hline\hline \\
   Dataset & n$^{a}_{H}$ & T$^{b}_{max}$  & kT$^{c}_{plasma}$  & $\Gamma$\tablenotemark{d}  & Flux\tablenotemark{e}$_{0.3-10.0 keV}$  & Flux\tablenotemark{f}$_{Disk}$ & $\chi^2$/dof \\
	\\
    \hline \\
2006         & 14.21$^{+0.87}_{-0.84}$& 0.141$^{+0.005}_{-0.005}$ & 1.00$^{+0.04}_{-0.04}$  & 2.66$^{+0.04}_{-0.04}$  & 4.03$^{+0.17}_{-0.39}$$\times$10$^{-12}$ & 1.74$^{+0.14}_{-0.28}$$\times$10$^{-12}$ & 431.07/265\\
\\
2008	& 14.67$^{+1.61}_{-1.48}$& 0.136$^{+0.010}_{-0.009}$ & 0.95$^{+0.06}_{-0.06}$  & 2.59$^{+0.06}_{-0.06}$  & 3.83$^{+0.24}_{-0.68}$$\times$10$^{-12}$ & 1.56$^{+0.24}_{-0.66}$$\times$10$^{-12}$ & 227.40/210 \\ 
\\
2010A	& 14.41$^{+1.01}_{-0.96}$& 0.144$^{+0.007}_{-0.006}$ & 1.01$^{+0.04}_{-0.04}$  & 2.58$^{+0.04}_{-0.04}$  & 4.55$^{+0.22}_{-0.43}$$\times$10$^{-12}$ & 1.72$^{+0.08}_{-0.35}$$\times$10$^{-12}$ & 403.69/277  \\ 
\\
2010B	& 13.04$^{+0.84}_{-0.80}$& 0.154$^{+0.007}_{-0.007}$ & 0.99$^{+0.05}_{-0.05}$  & 2.58$^{+0.04}_{-0.04}$  & 4.05$^{+0.15}_{-0.33}$$\times$10$^{-12}$ & 1.33$^{+0.11}_{-0.25}$$\times$10$^{-12}$ & 429.91/288  \\
\\
2011A 	&  14.58$^{+0.94}_{-0.90}$& 0.146$^{+0.007}_{-0.006}$ & 0.95$^{+0.06}_{-0.07}$  & 2.63$^{+0.04}_{-0.04}$  & 4.22$^{+0.19}_{-0.37}$$\times$10$^{-12}$ & 1.59$^{+0.17}_{-0.33}$$\times$10$^{-12}$ & 430.67/267 \\  
\\
2011B	&  13.44$^{+1.04}_{-0.99}$& 0.154$^{+0.009}_{-0.008}$ & 1.01$^{+0.07}_{-0.07}$  & 2.55$^{+0.05}_{-0.05}$  & 3.84$^{+0.16}_{-0.38}$$\times$10$^{-12}$ & 1.41$^{+0.15}_{-0.36}$$\times$10$^{-12}$ & 326.50/256 \\
\\  
 %\hline\hline \\ 
 %\end{tabular}\\
%\hspace{20cm} \begin{tabular}[h!]{lcccccccc} 
%\\
    \hline\hline \\
& & & & {\it tabs$\ast$(bmc+apec)} & & & \\
\\
    \hline\hline \\
 Dataset & n$_{H}$  & kT\tablenotemark{g}$_{disk}$  & $\alpha$=$\Gamma$\tablenotemark{h}-1  & $f$\tablenotemark{i}  & N\tablenotemark{j}$_{BMC}$ & kT$^{c}_{plasma}$ & $\chi^2$/dof \\
	\\
    \hline \\
2006         & 9.47$^{+1.04}_{-1.00}$ & 0.116$^{+0.004}_{-0.004}$ & 1.68$^{+0.04}_{-0.04}$ & 0.40$^{+0.05}_{-0.05}$ & 3.12$^{+0.31}_{-0.27}$$\times$10$^{-5}$ & 1.03$^{+0.04}_{-0.04}$ & 342.95/265 \\
\\
2008        & 10.79$^{+1.89}_{-1.78}$ & 0.110$^{+0.007}_{-0.006}$ & 1.59$^{+0.06}_{-0.06}$ & 0.39$^{+0.10}_{-0.10}$ & 3.10$^{+0.63}_{-0.47}$$\times$10$^{-5}$ & 0.97$^{+0.05}_{-0.05}$ & 210.30/210 \\	
\\
2010A	    & 9.73$^{+1.26}_{-1.22}$ & 0.116$^{+0.005}_{-0.005}$ & 1.59$^{+0.04}_{-0.04}$ & 0.45$^{+0.08}_{-0.08}$ & 3.35$^{+0.40}_{-0.33}$$\times$10$^{-5}$ & 1.03$^{+0.04}_{-0.04}$ & 359.92/277 \\	
\\
2010B	    & 7.90$^{+1.09}_{-1.05}$ & 0.121$^{+0.005}_{-0.005}$ & 1.60$^{+0.03}_{-0.03}$ & 0.52$^{+0.09}_{-0.08}$ & 2.75$^{+0.26}_{-0.22}$$\times$10$^{-5}$ & 1.03$^{+0.05}_{-0.04}$ & 364.36/288 \\	
\\
2011A       & 9.61$^{+1.16}_{-1.11}$ & 0.118$^{+0.005}_{-0.005}$ & 1.64$^{+0.04}_{-0.04}$ & 0.46$^{+0.07}_{-0.07}$ & 3.07$^{+0.33}_{-0.28}$$\times$10$^{-5}$ & 0.99$^{+0.05}_{-0.05}$ & 361.46/267 \\
\\
2011B	    & 8.69$^{+1.32}_{-1.25}$ & 0.121$^{+0.006}_{-0.006}$ & 1.57$^{+0.04}_{-0.04}$ & 0.47$^{+0.08}_{-0.08}$ & 2.74$^{+0.33}_{-0.26}$$\times$10$^{-5}$ & 1.05$^{+0.09}_{-0.06}$ & 299.37/256 \\
\\
    \hline\hline 
    \end{tabular}\\
\begin{flushleft}
\tablenotemark{a}{Total column density of hydrogen along the line of sight including the Galactic extinction (in units of 10$^{20}$cm$^{-2}$). We used the {\it tbabs} model in XSPEC.}
\tablenotemark{b}{Accretion disk temperature in keV. We used the {\it diskpn} model in XSPEC. The inner radius of the disk was fixed at 6GM/c$^{2}$.}
\tablenotemark{c}{The temperature of the surrounding plasma in keV. We used the {\it apec} model in XSPEC. The abundances were fixed at the solar value.}
\tablenotemark{d}{The photon index of the power law.}
\tablenotemark{e}{The total unabsorbed X-ray flux (in units of ergs cm$^{-2}$ s$^{-1}$) in the energy range of 0.3-10.0 keV.}
\tablenotemark{f}{The disk contribution to the total X-ray flux (in units of ergs cm$^{-2}$ s$^{-1}$) in the energy range of 0.3-10.0 keV.}
\tablenotemark{g}{The color temperature of the disk blackbody spectrum (in keV).}
\tablenotemark{h}{Spectral index of the power law portion of the energy spectrum.}
\tablenotemark{i}{The Comptonized fraction: fraction of the input blackbody photons that are Comptonized by the bulk motion of the in-falling material.}
\tablenotemark{j}{The normalization of the disk blackbody spectrum (in units of (L/10$^{39}$ erg s$^{-1}$)/(d/10 kpc)$^{2}$). This is an indicator of the disk flux.}
\end{flushleft}
}
\end{table*}

%%%%%%%%%%%%%%%%%%%%%%%%%%%%%%%%%%%%%%%%%%%%%%%%%%%

\section{Energy spectral Analysis}

We fit the X-ray spectra of NGC 5408 X-1 using the XSPEC (Arnaud 1996)
spectral fitting package and the EPIC response files were generated
using the $arfgen$ and $rmfgen$ tools which are part of the {\it
XMM-Newton} Science Analysis System (SAS) software.  Since, a primary
goal of the present work is to search for timing - spectral
correlations similar to those seen in accreting StBHs, we began our
analysis by characterizing the energy spectra using the same
phenomenological models often used to describe the X-ray spectra of
accreting StBHs.  In terms of XSPEC models, we used {\it diskpn +
power-law}.  We also explored the {\it bmc} model, Comptonization by
bulk motion (Titarchuk et al. 1997), because it has been used to derive
mass estimates from QPO frequency scaling arguments, and reference
spectral - timing correlations have been derived for a significant
sample of StBHs using the parameters derived from this model (see
Shaposhnikov \& Titarchuk 2009).  We start by describing some of the
specifics of our data extraction and reduction methods.

%%%%%%%%%%%%%%%%%%%%%%%%%%%%%%%%%%%%%%%%%%%%%%%%%%%
%%%%%%%%%%%%%%%%%%%%%%%%%%%%%%%%%%%%%%%%%%%%%%%%%%%

%\oddsidemargin=-0.9cm
%\vspace{-0.75cm}
\begin{table*}
%\oddsidemargin=-1.9cm
 % \begin{flushleft}
  \caption{ Summary of the energy spectral modeling of NGC 5408 X-1. Best-fitting parameters for the {\it tbabs$\ast$(diskpn+apec+cutoffpl)} model are shown. }\label{Table5}

{\scriptsize
\begin{center}
    \begin{tabular}{lccccccc}
     %\begin{tabular*}{20cm}[t]{lccccccc}
    \hline\hline \\	
& & &  {\it tbabs$\ast$(diskpn+apec+cutoffpl)} & & & \\
\\
    \hline\hline \\
   Dataset & n$^{a}_{H}$ & T$^{b}_{max}$  & kT$^{c}_{plasma}$  & $\Gamma$\tablenotemark{d}  & E$^{e}_{roll over}$  & $\chi^2$/dof \\
	\\
    \hline \\
2006         & 12.03$^{+0.97}_{-0.94}$& 0.153$^{+0.006}_{-0.006}$ & 1.03$^{+0.05}_{-0.04}$  & 1.88$^{+0.20}_{-0.21}$  & 3.80$^{+1.23}_{-0.79}$ & 375.48/264\\
\\
2008	& 12.93$^{+1.78}_{-1.66}$& 0.146$^{+0.011}_{-0.010}$ & 0.97$^{+0.06}_{-0.06}$  & 1.92$^{+0.33}_{-0.37}$  & 4.29$^{+3.96}_{-1.51}$ & 214.02/209 \\ 
\\
2010A	& 11.96$^{+1.15}_{-1.10}$& 0.160$^{+0.008}_{-0.008}$ & 1.04$^{+0.05}_{-0.05}$  & 1.84$^{+0.20}_{-0.22}$  & 4.18$^{+1.52}_{-0.93}$& 357.17/276  \\ 
\\
2010B	& 10.53$^{+1.03}_{-0.99}$& 0.171$^{+0.009}_{-0.008}$ & 1.02$^{+0.06}_{-0.06}$  & 1.92$^{+0.18}_{-0.20}$  & 4.92$^{+1.83}_{-1.11}$ & 385.27/287  \\
\\
2011A 	&  12.30$^{+1.09}_{-1.05}$& 0.160$^{+0.008}_{-0.008}$ & 0.98$^{+0.06}_{-0.07}$  & 1.92$^{+0.21}_{-0.23}$  & 4.28$^{+1.70}_{-1.01}$ & 389.94/266 \\  
\\
2011B	&  11.54$^{+1.24}_{-1.18}$& 0.167$^{+0.010}_{-0.009}$ & 1.05$^{+0.10}_{-0.08}$  & 1.98$^{+0.24}_{-0.26}$  & 5.56$^{+3.72}_{-1.69}$ & 307.22/255 \\
\\  
 \hline\hline \\ 

    \end{tabular}
  \end{center}
\tablenotemark{a}{Total column density of hydrogen along the line of sight including the Galactic extinction (in units of 10$^{20}$cm$^{-2}$). We used the {\it tbabs} model in XSPEC.}
\tablenotemark{b}{Accretion disk temperature in keV. We used the {\it diskpn} model in XSPEC. The inner radius of the disk was fixed at 6GM/c$^{2}$.}
\tablenotemark{c}{The temperature of the surrounding plasma in keV. We used the {\it apec} model in XSPEC. The abundances were fixed at the solar value.}
\tablenotemark{d}{The photon index of the cutoff power law. }
\tablenotemark{e}{The energy at which the exponential rollover of the spectrum occurs.}
}
\end{table*}

%%%%%%%%%%%%%%%%%%%%%%%%%%%%%%%%%%%%%%%%%%%%%%%%%%%

Since we are interested in exploring the correlations between the
timing and the spectral behavior, we elected to extract energy spectra
from the same time intervals as those used to extract the PDS (see the
first row of Tables 1 \& 2). Such synchronous measurements will tend
to minimize any offsets that could be induced by variations in the
system properties (both spectral and timing) within a given
observation. In addition to the standard filters described in \S 2,
the {\it (FLAG==0)} filter was imposed to get the highest quality
spectra. The combined ({\it pn+MOS}) average count rate of the source
varied in the range 1.19 - 1.46 cts s$^{-1}$ over all the
observations. Given such relatively low count rates, source pileup was
not an issue in any of the six observations. Since we are using both
the {\it pn} and the {\it MOS} data, we individually reduce the energy
spectra from each of the detectors using the same good time
intervals. This gave us three energy spectra ({\it pn, MOS1, MOS2})
for each of the six observations. A given model was fit simultaneously
to all the three spectra to derive the tightest constraints on the
best-fit parameters.  The {\it pn} and the {\it MOS} spectra were
binned to 1/3 of the FWHM of the pn and the MOS spectral resolution,
respectively.  We used the SAS task {\it specgroup} for this
purpose. Given the high number of total counts ($\approx$
(6-12)$\times$10,000), this gave us high-quality spectra in each case.

We began by fitting the spectra with a multi-colored disk ({\it
diskpn} in XSPEC, Gierlinski et al. 1999) + power law ({\it powerlaw}
in XSPEC) + an X-ray hot plasma ({\it apec} in XSPEC) as in S09. We
used the Tuebingen-Boulder ISM absorption model ({\it tbabs} in XSPEC)
to account for interstellar absorption. The hydrogen column density
was set as a free parameter. This model is very similar to that often
used to model the spectra of StBHs. Indeed, it is common to use this
or similar phenomenological models to determine the spectral state of
a StBH. However, here in addition to a thermal disk and a corona,
there is an X-ray emitting plasma in the model. This was realized by
S07 after identifying systematic emission-like features in the
residuals of the {\it diskpn + power-law} fit to the 2006 data. We
confirm their result that the $\chi^2$ improves significantly (by
$\sim$ 40-100 over all the six observations) with addition of the {\it
apec} component to the standard {\it multi-colored disk + power-law}
model. Such a feature is not exclusive to the ULX in NGC 5408, as NGC
7424 (Soria et al. 2006c) and Holmberg II X-1 (Dewangan et al. 2004)
also show evidence for the presence of an X-ray emitting plasma. In
addition, the high-resolution VLA radio observations at 4.9 GHz of the
counterpart of the X-ray source in NGC 5408 X-1 revealed a radio
nebula of $\approx$ 40 pc extent (Lang et al. 2007). Based on the
value of the power-law index of the radio spectrum, it was suggested
that the radio emission is likely from an optically thin synchrotron
emitting gas. Given this, it is possible that the putative X-ray hot
plasma is coincident with the radio nebula. The best-fitting
parameters of the above model for each of the six observations are
shown in the top panel of Table 4. We note that the X-ray temperature
of the plasma has remained constant across the suite of our {\it
XMM-Newton} observations.
%%%%%%%%%%%%%%%%%%%%%%%%%%%%%%%%%%%%%%%%%%%%%%%%%%%
\begin{figure}[ht!]

%\centering
%\vspace{-0.75cm}
\begin{center}
\includegraphics[width=3.75in, height=4.65in, angle=0]{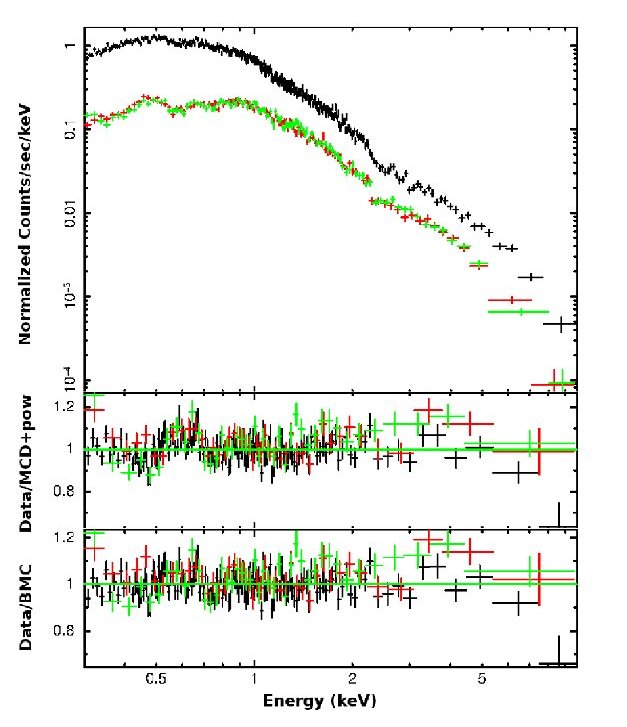}
\end{center}
%\begin{flushleft}
{\small {\textbf{Figure 4:} Top Panel: The X-ray energy spectrum
(0.3-10.0 keV) of NGC 5408 X-1 derived from the 2010B observation
(Table 1). The data from all the three detectors is shown. Black
corresponds to EPIC-pn, while green and red represent data from the
MOS1 and MOS2 detectors, respectively. The spectrum was re-binned for
clarity to have a significant detection of at least
10$\sigma$. However, no more than 10 neighboring bins were combined to
achieve this level of significance, i.e., {\it setplot rebin 10 10} in
XSPEC. Middle Panel: The ratio of the data to the model
defined by {\it tbabs$\ast$(diskpn + apec + powerlaw)} is shown.
Bottom Panel: The ratio of the data to the model defined by {\it
tabs$\ast$(bmc+apec)} is shown. Both the ratios were re-binned to have
a detection significance of at least 20 with no more than 20
neighboring bins combined, i.e., {\it setplot rebin 20 20} in XSPEC.}}
%\end{flushleft}
\label{fig:figure1a}
\end{figure}

%%%%%%%%%%%%%%%%%%%%%%%%%%%%%%%%%%%%%%%%%%%%%%%%%%%

We also fit the spectra with a model based on Componization by bulk
motion (Laurent \& Titarchuk 1999, {\it bmc} in XSPEC). As mentioned
above, this model has been used within the context of quantifying
timing - spectral correlations in StBHs (Shaposhnikov \& Titarchuk
2009). The underlying physical basis of this model is similar to that
of the multi-colored disk + power-law model, i.e., the presence of a
thermal disk and a hot electron corona. However, in this case the
power-law is produced by the Comptonization of soft photons (from the
disk) within a converging inflow onto the black hole.  In this model,
the normalization parameter derived from the fit mimics the disk flux
from the {\it diskpn} component in the other model we have
employed. We also include the {\it apec} component in this case. The
best-fitting model parameters including the absorbing hydrogen column
density using the {\it tbabs} in XSPEC for each of the six datasets are
shown in the bottom panel of Table 4.

It can be noted straight away that, for a given observation, the
plasma temperatures ({\it apec}) found from fitting the two spectral
models are consistent with each other (within the error bars). The
reduced $\chi^2$ values for the {\it diskpn+powlaw+apec} and the {\it
bmc+apec} model are in the range of 1.0-1.6 and 1.0-1.3 ($\sim$
210-290 degrees of freedom: last column of Table 4),
respectively. Clearly, they both give acceptable values of reduced
$\chi^2$. The latter model ({\it bmc+apec}) fits the data slightly
better (see Table 4) than the simple {\it diskpn+apec+powlaw}
model. However, the improvement in $\chi^2$ is not statistically
significant in all the cases and therefore it is not possible to rule
out one model over the other based on the reduced $\chi^2$ alone.  A
sample energy spectrum (using the 2010B data) is shown in Figure
4. Also shown in the figure are the ratios of the data to the folded
model in the two cases ({\it diskpn+apec+powlaw} and {\it
bmc+apec}). We use the spectral parameters derived here and the timing
properties from the previous section to search for correlations
between the two. The results of which are presented in the next
section.

It has been recently suggested (Gladstone et al. 2009) that the
high-quality X-ray spectra of ULXs can be better characterized by
complex comptonization models that predict a rollover at higher
energies ($\ga$ 3 keV).  A detailed study of the validity of such
models in the case of NGC 5408 X-1 is beyond the scope of the present
work, however, for the sake of completeness we also fit all our energy
spectra with a cut-off power law component rather than just the simple
power law model. In XSPEC we used the model, {\it
tbabs*(diskpn+apec+cutoffpl)}.  (the {\it cutoffpl} model describes a
power law with a high energy exponential rollover.)  The best fitting
parameters for this model are shown in Table 5. We note that this
model leads to a significant improvement in the $\chi^2$ compared to
the {\it tbabs*(diskpn+apec+pow)} model (compare the last column of
Table 5 with the last column of Table 4). We also searched for
possible timing-spectral correlations within the context of the {\it
cutoff powerlaw} model, viz., QPO frequency vs the cutoff energy.  We
do not detect any clear correlation between the QPO frequency and the
rollover energy. This is mainly due to poor statistics at the high end
($\ga$ 5 keV) of the energy spectra which result in large
uncertainties on the rollover energies.  

The measurement of fluorescent lines from elements like Fe, Ni, Cr, Ca
etc. is another important probe of the vicinity of the X-ray emitting
region. In principle, these reflection features can be produced by
hard X-rays, presumably from the Comptonizing corona, irradiating cold
material, e.g., in the disk. Among all such emission lines with their
rest-frame energies in the {\it XMM-Newton} bandpass (0.2-10.0 keV),
the iron feature is the most prominent (e.g., Matt et al. 1997). Given
the subtle changes in the energy spectra over the suite of
observations (see Table 4), it does not seem unreasonable to combine
all the data while searching for weak emission features. We searched
for iron emission line features in the energy range of 6.0-7.0 keV,
but did not detect any obvious features. Using an effective exposure
of $\approx$ 540 ks (all datasets, see Table 1), we were able to place
tight constraints on the equivalent width of any feature in the energy
range of 6.0-7.0 keV. Assuming an unresolved narrow feature of width
10 eV, the upper limit (90\% confidence) on the equivalent width of an
emission line at 6.4 keV is 5.4 eV. We then check for the possibility
of broad emission lines in the energy range 6.0-7.0 keV. The upper
limits on the equivalent width assuming a broad emission line of width
0.3 keV and 1.0 keV are 11.6 eV and 11.0 eV, respectively.

The apparent weakness of iron emission from the disk can be due to a
number of factors: an accretion disk that is completely ionized,
scattering of the reflected iron-line photons in an optically thick
corona (Matt et al. 1997), especially when the corona is a thin layer
above the accretion disk or a low iron abundance in the accretion disk
(Matt et al. 1997). Further, a high inclination angle for the
accretion disk combined with one or more of the above factors can
decrease the intensity of the iron-line. Recently, Gladstone et
al. (2009) have analyzed energy spectra of a sample of ULXs (including
NGC 5408 X-1) and have suggested that these sources might be operating
in a new accretion state, with rollover at higher energies ($>$ 2.0
keV) as a characteristic signature of such a state. Thus, the nature
of the spectrum might be such that there are not enough photons at
higher energies to generate a detectable iron-line fluorescent feature
(Cole Miller, private communication).

%%%%%%%%%%%%%%%%%%%%%%%%%%%%%%%%%%%%%%%%%%%%%%%%%%%
%%%%%%%%%%%%%%%%%%%%%%%%%%%%%%%%%%%%%%%%%%%%%%%%%%%

\begin{table*}
  \begin{center}
  \caption{{Comparing the PDS and the energy spectral properties of NGC 5408 X-1 with the steep power-law (SPL) state in stellar-mass black holes}}\label{Table6} \centering
{\footnotesize
    \begin{tabular}{lcc}
    \hline\hline \\
   X-ray State & Definition\tablenotemark{a} \\
	\\
    \hline \\
	   			 & Presence of power-law component with $\Gamma$ $>$ 2.4 \\
Steep Power Law State		      & 	Power continuum: r$^{b}$ $<$ 0.15\\
(SPL) 	            &  Either disk fraction$^{c}$ $<$ 80\% and 0.1-30 Hz QPOs present with rms amplitude $>$ 0.01\\
			     &	    or disk fraction$^{c}$ $<$ 50\% with no QPOs present\\
\\
\hline
\\
X-ray State & {\bf Data from NGC 5408 X-1\tablenotemark{d} }   \\
	\\
    \hline \\
	   			 &		 Presence of power-law component with $\Gamma$ in the range 2.5-2.7 \\
Steep Power Law State ?	      & 	Power continuum: r$^{\ast}$ in the range $\approx$ 0.36 - 0.43\\
(SPL) 	             &  Disk fraction$^{e}$ in the range $\sim$ 30-50\% and\\
				&   10-40 mHz QPOs present with rms amplitude in the range 0.10-0.45\\
					      \\
\hline

    \hline\hline
    \end{tabular}
}
	\end{center}
{\footnotesize
\tablenotemark{a}{In case of stellar-mass black holes, the definition is using data in the energy range of 2-20 keV.}\\
\tablenotemark{b}{Total rms power integrated over 0.1-10.0 Hz.}\\
\tablenotemark{c}{Fraction of the total 2-20 keV unabsorbed flux.}\\
\tablenotemark{d}{In case of NGC 5408 X-1, we use the {\it XMM-Newton} bandpass of 0.3-10.0 keV.}\\
\tablenotemark{$\ast$}{We integrate the total power (continuum + QPOs) in the frequency range of 0.001 - 0.5 Hz (same as the frequency range used for power spectra fitting). The power spectra were derived using all the photons in the energy range of 1.0-10.0 keV (see Table 3 for the parameters).}\\
\tablenotemark{e}{Fraction of the total 0.3-10 keV unabsorbed flux.}\\
\tablenotemark{f}{Fraction of the total 0.3-10 keV unabsorbed flux.}\\
}

\end{table*}
%%%%%%%%%%%%%%%%%%%%%%%%%

\section{Timing - Spectral Correlations}

A primary goal of the new observations of NGC 5408 X-1 is to further
test the preliminary classification of the mHz QPOs detected from NGC
5408 X-1 as the analogs of the 0.1-15 Hz Low-Frequency, type-C QPOs
(LFQPOs) detected from accreting StBHs. If a strong connection can be
demonstrated, then scaling of the characteristic frequencies (e.g.,
QPO centroid frequency) can be better justified to estimate the mass
of the black hole within this source.

As mentioned earlier, scaling relations using the PDS break frequency
have been successfully tested using power spectra of both stellar and
super-massive black holes of known mass (McHardy et al. 2006 and
K\"ording et al. 2007). Based on the qualitative nature of the PDS and
the energy spectra derived using the 2006 and 2008 data, S09 suggested
that the mHz QPOs from NGC 5408 X-1 may indeed be LFQPO analogs. More
specifically, they proposed that its X-ray state was analogous to that
of a StBH in the so-called steep power-law (SPL) state exhibiting
type-C QPOs, but at the same time emitting a few$\times$10 higher
X-ray flux. Having analyzed the data from all six observations (\S 3
\& 4) we confirm that the derived power- and energy-spectral
parameters are qualitatively consistent with the source being in a SPL
state exhibiting type-C QPOs (based on the state descriptions in
McClintock \& Remillard 2006). A summary of the working definition of
the SPL state with type-C QPOs and the PDS \& energy spectral
parameters of NGC 5408 X-1 are given in Table 6.

A further step in understanding the nature of these QPOs is to
determine if they show the same evolutionary behavior in the QPO
parameters as those exhibited by the StBH LFQPOs. One such
characteristic behavior is the QPO rms amplitude versus centroid
frequency relationship. Our results for this relation from NGC 5408
X-1 are shown in the left and right panels of Figure 3 in two
different energy bands (1.0-10.0 keV and 0.5-5.0 keV, respectively).
Comparing these plots with those from StBHs, as in Figure 11 of
McClintock et al. (2009), the results appear to be at least
qualitatively consistent with the behavior of SPL data from some StBHs
(see for example, the green triangles in Figure 11 of McClintock et
al. 2009).  However, because a tight correlation appears to break down
in the SPL state it is hard to conclude definitively that the NGC 5408
X-1 behavior is exactly analogous.  
 
%%%%%%%%%%%%%%%%%%%%%%%%%%%%%%%%%%%%%%%%%%%%%%%%%%%
%%%%%%%%%%%%%%%%%%%%%%%%%%%%%%%%%%%%%%%%%%%%%%%%%%%
\begin{figure*}

\begin{center}
\includegraphics[width=5in, height=5in, angle=0]{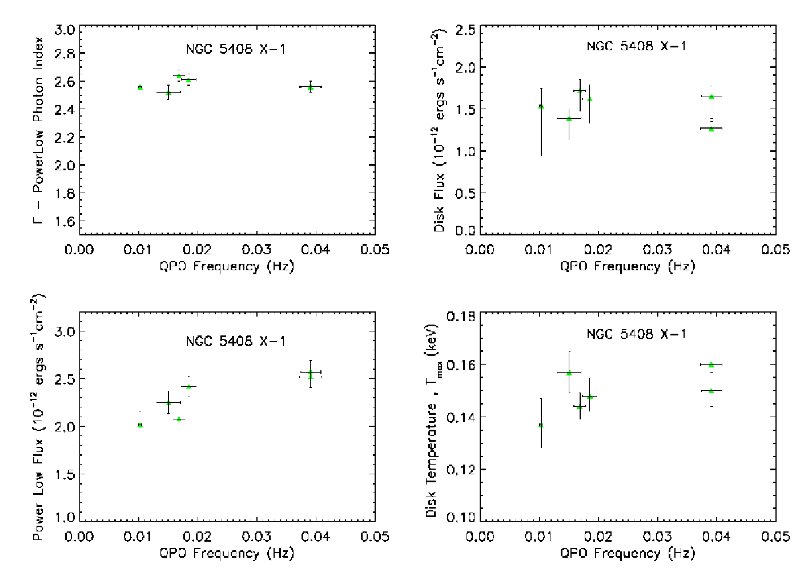}
\end{center}
{\small {\textbf{Figure 5A:} Timing-Spectral Correlations using the
phenomenological model: diskpn + powerlaw. Six observations/data
points were used to search for timing-spectral correlations. Top
Left Panel: The photon index of the power-law (Y-axis) is plotted
against the centroid frequency of the QPO (X-axis). Top Right
Panel: The disk contribution to the total flux (Y-axis) in the
energy range of 0.3-10.0 keV is plotted against the centroid frequency
of the QPO. Bottom Left Panel: Flux from the power-law
component (Y-axis) is plotted against the QPO centroid frequency
(X-axis). Bottom Right Panel: For the sake of completeness,
the disk temperature (Y-axis) is plotted against the QPO centroid
frequency (X-axis). Within the present scheme of X-ray states in
StBHs, these results are qualitatively consistent with NGC 5408 X-1
being in the steep power law state exhibiting type-C QPOs, with the
characteristic timescales scaled down by a factor of $\approx$ a
few$\times$10.}}
\label{fig:Figure 5A}
\end{figure*}

%%%%%%%%%%%%%%%%%%%%%%%%%%%%%%%%%%%%%%%%%%%%%%%%%%%
%%%%%%%%%%%%%%%%%%%%%%%%%%%%%%%%%%%%%%%%%%%%%%%%%%%

Another important signature of the Type-C LFQPOs in StBHs is their
dependence on the energy spectral features, i.e., the timing -
spectral correlations. Using the X-ray state classifications as in
McClintock et al. (2006), LFQPOs in StBHs are usually detected in the
low/hard state, SPL state and during the transition between these two
states. Within the context of the multi-colored disk + power-law model
parameters, it is known that the centroid frequency of these LFQPOs is
correlated with the disk flux and the photon index of the power-law
component. In StBHs, the typical behavior is that the QPO centroid
frequency is positively correlated with the disk flux and the photon
index of the power law. At a certain higher QPO centroid frequency
($\approx$ 5-10 Hz) the relationship seems to flatten (saturate) and
even appears to start reversing in some cases (Vignarca et
al. 2003). Furthermore, the correlation is not exactly the same for
all StBHs, i.e., the value of the slope of the correlation and the
photon index/disk flux at saturation are different for different
sources and can be different for the same source in a different
outburst (Vignarca et al. 2003; Shaposhnikov \& Titarchuk 2009).  In
Figure 5 we show plots of several derived spectral parameters and
fluxes (using the {\it diskpn} + {\it power-law} fits) as a function
of the QPO centroid frequency. These can be compared with similar
plots derived from StBH systems (see, for example, Figures 8 \& 9 in
McClintock et al. 2009; Figure 10 of Vignarca et al. 2003). Again,
there appears to be qualitative consistency between the behavior
exhibited by NGC 5408 X-1 and StBHs, but the full range of behavior is
not yet seen, again because the range of spectral variations in NGC
5408 X-1 (most notably in the power-law index) is too modest. Additional
measurements of QPO properties at lower values of the power-law photon
index, for example, could provide a more definitive test.

%%%%%%%%%%%%%%%%%%%%%%%%%%%%%%%%%%%%%%%%%%%%%%%%%%%

\begin{figure*}

\begin{center}
%\vspace{-0.75cm}
\includegraphics[width=6.5in, height=4.33in, angle=0]{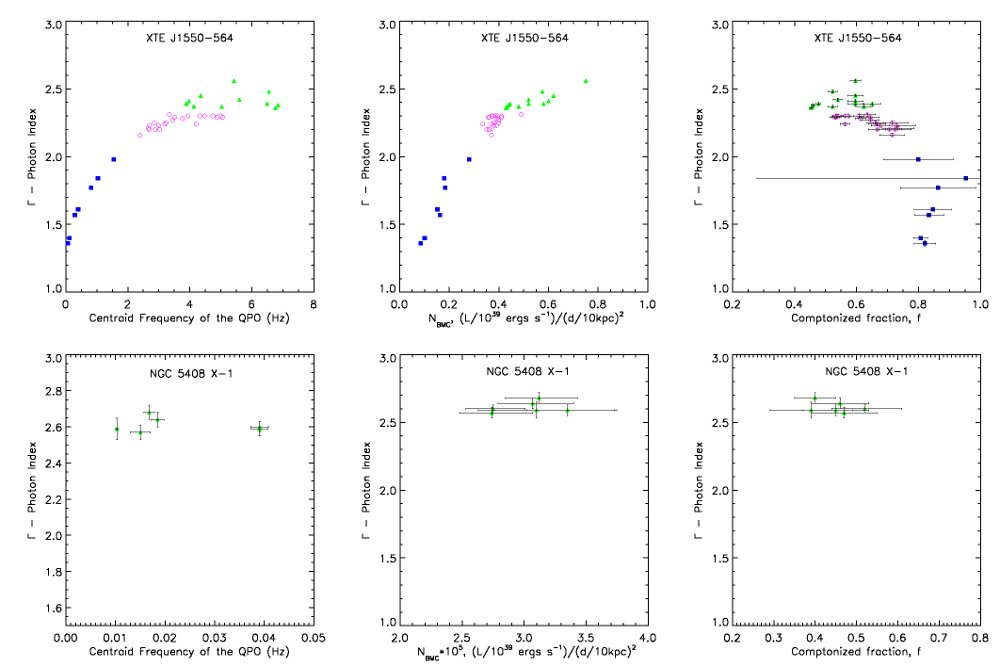}
\end{center}
{\small {\textbf{Figure 5B:} Timing-Spectral Correlations using the BMC model.
  Top Panels: The behavior of a typical stellar-mass black
  hole (1998 outburst of StBH XTE J1550-564) is shown (we plot data from Shaposhnikov \& Titarchuk
  2009). For clarity, the X-ray state of the source is highlighted
  using the following color scheme. Blue squares indicate the low/hard
  state, magenta corresponds to a state intermediate between the
  low/hard and the SPL state while the green triangles highlight the
  SPL state. From left to right, the dependence of the centroid
  frequency of the QPO, BMC normalization (disk flux) and the
  Comptonized fraction on the power law index are shown. The error
  bars are not indicated when their size is smaller than the size of
  the data point. Bottom Panel: Similar plots using the data
  from NGC 5408 X-1. The results from the ULX NGC 5408 X-1 are
  qualitatively consistent with the source being in an X-ray state
  very similar to the steep power-law state in stellar-mass black
  holes (Compare with the green triangles on the top panel).}}
\label{fig:figure1a}
\end{figure*}

%%%%%%%%%%%%%%%%%%%%%%%%%%%%%%%%%%%%%%%%%%%%%%%%%%%
%%%%%%%%%%%%%%%%%%%%%%%%%%%%%%%%%%%%%%%%%%%%%%%%%%%
%%%%%%%%%%%%%%%%%%%%%%%%%%%%%%%%%%%%%%%%%%%%%%%%%%%

One can also investigate the spectral - temporal correlations in the
context of the spectral parameters derived using the {\it bmc} model.
Indeed, Shaposhnikov \& Titarchuk (2009) explored such correlations in
detail for a sample of StBHs using RXTE data.  They also developed a
set of fitting functions to quantify the observed correlations and
used these to obtain mass estimates by scaling arguments.
Figure 5B compares the results from NGC 5408 X-1 with those from StBHs
using the {\it bmc} spectral fits. We used the published data from
Shaposhnikov \& Titarchuk (2009) to produce representative plots for a
StBH (XTE J1550-564, top panels). Shown from left to right are the QPO
centroid frequency versus the power-law photon index ($\Gamma$), the
{\it bmc} normalization (disk flux) versus $\Gamma$ and the
Comptonized fraction, $f$ versus $\Gamma$, respectively. The data
points are color coded to highlight the low/hard state (blue) and the
SPL state (green). We used the data from the 1998 outburst of XTE
J1550-564 which clearly demonstrates the typical behavior of StBHs
(top panels).  Comparisons of these plots with those in, for example,
Shaposhnikov \& Titarchuk (2009) lead to similar conclusions as with
the results in Figure 5A.  There is qualitative consistency with the
Type-C identification, but the classification is not definitive.  In
order to make a more secure association of the QPOs in NGC 5408 X-1
with Type-C (SPL state) QPOs in the StBHs, we need, at a minimum, to
obtain observations of the source over a greater range of power-law
spectral indices.

\section{Discussion}

It has been established that black hole masses scale with the break
frequency of their PDS (McHardy et al. 2006, K\"ording et al. 2007) 
and this relation is known to hold over 6 orders of magnitude in mass,
i.e., from StBH to super-massive black holes. Furthermore, at least in
StBHs, the break frequency of the PDS is known to strongly correlate
with the centroid frequency of the LFQPOs (Wijnands et
al. 1999). Therefore, it is reasonable to assume that the mass of the
black hole scales with the centroid frequency of the LFQPOs. This can
be especially useful in cases where black hole mass measurements are
otherwise difficult, viz., in ULXs. However, an important caveat is
to ensure that the QPOs used for comparison are similar in nature.

Our new results confirm the earlier result from S09 that the
qualitative nature of the PDS and the energy spectrum of NGC 5408 X-1
are very similar to that of StBHs in the SPL state (see Table 6). At
the same time, the characteristic timescales within the PDS are lower
by a factor of $\approx 100$, while the X-ray luminosity is higher by
a factor of a few$\times$10, when compared to a typical StBH in such a
state. However, based on the scaling between the break frequency of
the PDS and the centroid frequency of the LFQPOs (Centroid frequency
of LFQPOs $\approx$ 12$\times$break frequency of the PDS), the
association of the mHz QPOs from NGC 5408 X-1 with the LFQPOs from
StBHs has been questioned (Middleton et al. 2011)

A striking feature of the behavior of NGC 5408 X-1 is that the energy
spectrum of the source has remained roughly constant over the current
suite of observations, while the properties of the most prominent QPO
(e.g., rms amplitude, centroid frequency) have changed
significantly. These results appear consistent with the source being
in a spectral state similar to the SPL state in StBHs. More
specifically, in StBHs, certain spectral properties like the disk
flux, the photon index of the power-law correlate positively with the
QPO centroid frequency up to a certain frequency ($\approx$ 2-10 Hz)
beyond which the relationship turns around or remains roughly
constant. Detailed analysis of the timing - spectral correlations in
StBH sources such as XTE J1550-564 and H1743-322 has revealed a
comprehensive picture (McClintock et al. 2009) of their
behavior. Tracking these two sources as they evolved from the low/hard
state into the SPL state, it was realized that the spectral properties
like the disk flux and the power-law photon index are tightly
correlated with the QPO centroid frequency during the low/hard state
and during a transition phase intermediate between the low/hard and
the SPL state; and the relationships seem to saturate/breakdown
(spectral properties remain roughly constant with increase in the QPO
frequency) as the sources entered the SPL state.  It may be that we
are in fact seeing NGC 5408 X-1 in a similar state, that is, we may be
seeing the ``saturated'' portion of the relation. 

Given that the overall accretion timescales change with the mass of
the black hole (timescales increase as the mass increases from StBHs
to super-massive black holes, e.g., McHardy et al. 2006), it is
possible that we are seeing a similar timing - spectral behavior, and
that the apparent longevity of the source in the SPL state might also
be due to its having a higher mass. There may be a useful analogy to
AGN in that their spectra are known to remain roughly constant on
timescales of at least a few years (Markowitz et al. 2003) while
transient StBH systems undergo strong energy spectral changes on
timescales of the order of a few days (McClintock et al. 2006 and
references therein). It could be that the persistence of NGC 5408 X-1
in the approximately same spectral state is related to it being of
intermediate mass between StBHs and AGN.  Alternatively, perhaps the
source is in an accretion state, like the so-called ultraluminous
state of Gladstone et al. (2009), that is just not well sampled by the
behavior of Galactic StBHs. We note that the indication for a rollover
in the power-law at 4-5 keV (see \S 4 and Table 5) provides some
support for this interpretation, in which case the spectral state of
NGC 5408 X-1 may not be exactly analogous to the SPL state of StBHs.

However, either conclusion will remain tentative until NGC 5408 X-1 is
observed in a state with a lower value of the power-law photon index,
i.e., a state similar to the low/hard state in StBHs. Such an
observation would clearly be important in order to make a more
definitive identification of its QPOs in the context of those observed
in StBHs.

Assuming the constancy of the energy spectral features, i.e., disk
flux, $\Gamma$, with the increasing QPO centroid frequency (see Figure
5) is analogous to the saturation observed in the timing - spectral
correlations in StBHs, one can estimate the minimum mass of the black
hole in NGC 5408 X-1. If we are indeed seeing only the saturated
portion of the complete timing - spectral correlation curve, then the
minimum QPO frequency observed in NGC 5408 X-1 serves as an upper
limit on the so-called transition frequency, i.e., the frequency
beyond which the timing - spectral correlations tend to saturate. In
order to derive a mass estimate, this particular frequency can be
scaled to a reference StBH of known mass with a measured transition
frequency ($\nu_{trans}$). We can use the transition frequency as
defined by Shaposhnikov \& Titarchuk (2009) in their fits of QPO and
spectral parameters.  We find a sample of three StBHs with mass and
transition frequency measurements in Shaposhnikov \& Titarchuk (2009)
that are suitable for this purpose. Further, it is known that the same
source can exhibit different tracks (timing - spectral correlation
curves), i.e., different transition frequencies, in various outburst
episodes.  Given that we are interested in a lower limit on the mass
of the black hole in NGC 5408 X-1, we consider the lowest transition
frequency for a given source. The minimum mass of the black hole in
NGC 5408 X-1, can then be calculated as
$\nu_{trans}$$\times$Mass$_{Reference}$/(minimum QPO frequency). The
mass estimates using the three reference sources, GRO 1655-40, XTE
J1550-564 and GX 339-4 are reported in Table 7 ($\sim$ 1000
M$_{\odot}$), and show substantial overlap with those reported by S09.

%%%%%%%%%%%%%%%%%%%%%%%%%%%%%%%%%%%%%%%%%%%%%%%%%%%
%%%%%%%%%%%%%%%%%%%%%%%%%%%%%%%%%%%%%%%%%%%%%%%%%%%
%%%%%%%%%%%%%%%%%%%%%%%%%%%%%%%%%%%%%%%%%%%%%%%%%%%
%\oddsidemargin=-0.9cm

\begin{table}

    \caption{{ Mass estimate of the black hole in NGC 5408 X-1}}\label{Table7} 
{\scriptsize
\begin{center}
    \begin{tabular}[t]{lcccc}
     %\begin{tabular*}{20cm}[t]{lccccccc}
    \hline\hline \\
	  &  	 &				  & Minimum		&	\\
Reference & Mass & $\nu_{trans}$\tablenotemark{a} & Mass$^{b}_{5408}$ & References \\
	  & (M$_{\odot}$) & (Hz) 		      & (M$_{\odot}$)   &  			\\
	\\
    \hline \\
GRO J1655-40 & $6.3\pm0.5$ & $3.0\pm0.1$  & 1720 &  1, 4 \\

	\\

XTE J1550-564 & $9.5\pm1.1$ & $1.84\pm0.07$  & 1490 & 2, 4 \\

	\\

GX 339-4 & $>$ 6 & $1.4\pm0.2$ & 820 & 3, 4 \\
\\
    \hline\hline
    \end{tabular}
\end{center}
}
%\newline
\begin{flushleft}
\tablenotemark{a}{The transition frequencies were estimated by Shaposhnikov et al. (2009). The fitting functions for the timing-spectral correlations are described therein.}\\
\tablenotemark{b}{Minimum mass of the black hole in the ULX = ($\nu_{trans}$$\times$mass)$_{Reference StBH}$/10.28 mHz, where 10.28 mHz is the minimum QPO frequency detected from NGC 5408 X-1. We consider the error bars on the individual parameters and report the lower value of the mass. }\\
{\small {\bf References} - (1) Greene et al. 2001; (2) Orosz et al. 2002 (3) Mu{\~n}oz-Darias et al. 2008 (4) Shaposhnikov et al. 2009.}\\
\end{flushleft}
\end{table}

%%%%%%%%%%%%%%%%%%%%%%%%%%%%%%%%%%%%%%%%%%%%%%%%%%%
%%%%%%%%%%%%%%%%%%%%%%%%%%%%%%%%%%%%%%%%%%%%%%%%%%%
%%%%%%%%%%%%%%%%%%%%%%%%%%%%%%%%%%%%%%%%%%%%%%%%%%%

\section{Summary}

The ULX in the irregular dwarf galaxy, NGC 5408 has been suggested to
harbor an intermediate-mass black hole (Mass$_{BH}$ $\approx$ 100 - a
few$\times$1000 M$_{\odot}$) (S09). This mass estimate was strictly
based on the assumption that the mHz QPOs seen from this source are
the type-C analogs of the low-frequency QPOs (0.1-15 Hz) from StBHs.
Here we have presented results from new observations of NGC 5408 X-1
in which we searched for timing - spectral correlations similar to
those often exhibited by the type-C LFQPOs from StBHs. Our analysis of
multi-epoch {\it XMM-Newton} data from NGC 5408 X-1 reveals that
certain characteristic features of the power spectra, especially the
QPO centroid frequency, changed significantly. However, the energy
spectrum has remained roughly constant. These results can be
interpreted in two ways. This could be due to complete independence of
the timing properties on the energy spectra, unlike StBHs, in which
case mass scalings derived from the QPOs are likely to be problematic;
or there is in fact a correlation, but we are seeing only the
saturated part of the correlation behavior (constancy of the
$\Gamma$/disk flux with increasing QPO frequency). Such saturation is
often seen in StBHs (e.g., Vignarca et al. 2003). Assuming we are
seeing this saturated portion of the correlation curve, we estimate
the lower limit on the mass to be $\approx$ 800 M$_{\odot}$. At least
one observation with the source in a low/hard-like state ($\Gamma$ $<$
2.0) is necessary to resolve the issue of whether the timing and
spectral properties are correlated as in StBHs or not.

%%%%%%%%%%%%%%%%%%%%%%%%%%%%%%%%%%%%%%%%%%%%%%%%%%%%%%%%%%%%%%%%%
%%%%%%%%%%%%%%%%%%%%%%%%%%%%%%%%%%%%%%%%%%%%%%%%%%%%%%%%%%%%%%%%%
%\vfill\eject

%%%%%%%%%%%%%%%%%%%%%%%%%%%%%%%%%%%%%%%%%%%%%%%%%%%
%%%%%%%%%%%%%%%%%%%%%%%%%%%%%%%%%%%%%%%%%%%%%%%%%%%
%%%%%%%%%%%%%%%%%%%%%%%%%%%%%%%%%%%%%%%%%%%%%%%%%%%

\end{document}